\documentclass[11pt]{article}

\pdfoutput=1

\usepackage[textwidth=15.5cm,textheight=22.5cm]{geometry}
\usepackage{bm}
\usepackage{amsmath,amssymb,color,graphicx,amscd,amsfonts}
\usepackage{latexsym}
\usepackage{graphicx}
\usepackage{cite}
\usepackage{subfig}
\usepackage{bbm}
\usepackage{ulem}

\usepackage{datetime}

\begin{document}

\thispagestyle{empty}

\setcounter{page}{0}

\begin{flushright} 
 DMUS-MP-22/03
 \\
\today

\end{flushright} 

\vspace{0.1cm}

\begin{center}
{\Large Matrix Entanglement
}
\end{center}

\vspace{0.2cm}

\renewcommand{\thefootnote}{\alph{footnote}}
\begin{center}
Vaibhav Gautam$^a$, Masanori Hanada$^a$, Antal Jevicki$^b$ and Cheng Peng$^c$\\
\vspace{0.2cm}
$^a$\textit{Department of Mathematics, University of Surrey, Guildford, Surrey, GU2 7XH, UK}\\
$^b$\textit{Department of Physics, Brown University,182 Hope Street, Providence, RI 02912, USA}\\
$^c$\textit{Kavli Institute for Theoretical Sciences (KITS), \\
University of Chinese Academy of Sciences, Beijing 100190, China}\\
\end{center}

\begin{abstract}

In gauge/gravity duality, matrix degrees of freedom on the gauge theory side play important roles for the emergent geometry. In this paper, we discuss how the entanglement on the gravity side can be described as the entanglement between matrix degrees of freedom. 
Our approach, which we call `matrix entanglement', is different from `target-space entanglement' proposed and discussed recently by several groups. 
We consider several classes of quantum states to which our approach can play important roles.  
When applied to fuzzy sphere, matrix entanglement can be used to define the usual spatial entanglement in two-brane or five-brane world-volume theory nonperturbatively in a regularized setup. 
Another application is to a small black hole in AdS$_5\times$S$^5$ that can evaporate without being attached to a heat bath, 
for which our approach suggests a gauge theory origin of the Page curve.
The confined degrees of freedom in the partially-deconfined states play the important roles. 
\end{abstract}

\newpage

\tableofcontents

\section{Introduction}
The importance of quantum entanglement is getting more and more widely appreciated. 
In the context of quantum gravity, a proper understanding of quantum entanglement is important for the resolution of the black hole information puzzle. (See e.g., Refs.~\cite{Almheiri:2020cfm,Raju:2020smc} for recent reviews.) Furthermore, quantum entanglement might be a key to understanding the emergent space in holography~\cite{VanRaamsdonk:2010pw}. In this paper, we would like to introduce a new tool having such a big picture in mind. 

In the Ryu-Takayanagi prescription~\cite{Ryu:2006bv} for holographic entanglement entropy, quantum field theories with nonzero spatial dimensions are assumed. Then, the space of dual quantum field theory is split into two parts. Let us call entanglement defined in this way as `spatial entanglement'. Spatial entanglement is not the only way to define quantum entanglement. In principle, any partition of the Hilbert space is allowed. For example, when Yang-Mills theory is considered, matrix degrees of freedom (color degrees of freedom) can be partitioned into two or more pieces. 
The necessity of considering such a partition is obvious because the Ryu-Takayanagi prescription cannot be applied to an important class of models --- matrix models --- which do not have spatial dimensions. 
One might think it is merely another option, but in fact, it is natural to expect that such a partition is of crucial importance, presumably more important than the partition of the spatial dimensions, for the following reasons.  

A remarkable application of fine-grained entropy formulas \cite{Hubeny:2007xt,Lewkowycz:2013nqa,Faulkner:2013ana,Engelhardt:2014gca} obtained by refining or generalizing the Ryu-Takayanagi formula ~\cite{Ryu:2006bv} is the derivation of the Page curve~\cite{Page:1993wv} from gravity side based on the entanglement island proposal~\cite{Penington:2019npb,Almheiri:2019psf}. 
Interestingly, and a little bit frustratingly, such a derivation does not use microscopic details of quantum gravity. In fact, it is believed that these entropy formulas are generic features of quantum systems coupled to gravity, which does not require holography~\cite{Almheiri:2020cfm}.
To make progress in the understanding of quantum gravity, we have to learn how such formulas can be understood in terms of microscopic degrees of freedom. In the holographic approach, the matrix model can likely provide us with a dual description of the evaporating black hole~\cite{Banks:1997tn,Banks:1997cm,Horowitz:1997fr,Bergner:2021goh}, and hence, entanglement between matrix degrees of freedom should play the key role in such a direction of study. It would be natural to expect that the same situation whether we consider matrix model or QFT. 

Another, but deeply related, reason suggesting the importance of considering such a partition is that matrix degrees of freedom play important roles in emergent geometry. The story about black hole evaporation we mentioned in the previous paragraph can likely be understood as an example of emergent holographic geometry out of matrix degrees of freedom. A more established example is the emergence of higher-dimensional quantum field theories (e.g.,~3d maximal super Yang-Mills) from lower-dimensional theories (e.g.,~the BMN matrix model~\cite{Berenstein:2002jq}) on certain background. 
In this example, spatial degrees of freedom in the higher-dimensional theory is encoded into matrix degrees of freedom in the lower-dimensional theory. This is analogous to a lattice regularization with the large-$N$ limit corresponding to the continuum limit. Hence the spatial partition in the higher-dimensional theory is a special case of the matrix partition in the lower-dimensional theory.

In this paper, we will discuss how the entanglement between matrix degrees of freedom, which we call `matrix entanglement', can be defined and how it is related to dual gravity theory.
We do this first for the matrix model, and then give a generalization to QFT.   
We will apply our method to several examples including fuzzy spheres and a small black hole that can evaporate completely~\cite{Horowitz:1999uv}. 
Our `matrix entanglement' is different from `target space entanglement'~\cite{Mazenc:2019ety,Das:2020jhy,Das:2020xoa,Hampapura:2020hfg,Frenkel:2021yql} considered by several groups of researchers in the past, as explained in Sec.~\ref{sec:target-space-EE}.
We hope that our findings will convince the readers that the consideration of such matrix entanglement is of crucial importance, and without it, we cannot fully utilize the power of holographic entanglement formulas on the gravity side. 

In U($N$) super Yang-Mills theory or matrix model dual to superstring theory, dynamical variables are $N\times N$ matrices with $N^2$ color degrees of freedom. 
We treat $N^2$ colors similarly to `particles' and split them into two or more pieces. 
That we work at each fixed $N$ can be interpreted that we are dealing with the first-quantized system of $N$ D-branes.
Intuitively, we interpret that some D-branes form a bound state such as a black hole, other D-branes form other objects, and such objects interact with each other. We will consider entanglement between these objects. 

For such a definition of entanglement to make sense, we have to find a natural way of splitting matrix degrees of freedom into two or more pieces. Such a partition is obtained by understanding the meaning of `eigenvalues' in gauge theory properly~\cite{Hanada:2021ipb}. An important example is the partially-deconfined states dual to small black hole, in which an U($M$)-subgroup of U($N$) ($0<M<N$) is deconfined~\cite{Hanada:2016pwv,Berenstein:2018lrm,Hanada:2018zxn,Hanada:2019czd,Hanada:2020uvt}. 
The confined and deconfined sectors in the partially-deconfined states provide us with a natural partition, and gauge fixing can be regarded as a superselection~\cite{Hanada:2019czd}. 
One of our motivations is to define the entanglement between black hole and Hawking radiations solely in terms of dual QFT or matrix model without referring to gravity, as a tool to study quantum gravity based on the first principles. 

This paper is organized as follows.
In Sec.~\ref{sec:matrix-geometry}, we explain how the geometry on the gravity side can be encoded into matrix degrees of freedom in the matrix model or QFT. 
It is important to note the relationship between the gauge-invariant Hilbert space $\mathcal{H}_{\rm inv}$ and the extended Hilbert space $\mathcal{H}_{\rm ext}$ that contains gauge-non-singlet states, discussed in Sec.~\ref{sec:H_inv-vs-H_ext}. The latter is suitable for geometric interpretations. 
In Sec.~\ref{sec:wave-packet}, we use wave packets in the extended Hilbert space to encode geometry into matrix degrees of freedom. 
Based on such a geometric interpretation, the matrix entanglement is introduced in Sec.~\ref{sec:EE-definition}. 
For a class of quantum states we consider, geometric interpretation becomes clear after gauge fixing, and we can define the matrix entanglement in a straightforward manner. 
In Sec.~\ref{sec:fuzzy_sphere}, we consider the matrix entanglement associated with the fuzzy-sphere background of the BMN matrix model. In appropriate large-$N$ limits, 3d SYM or 6d $\mathcal{N}=(2,0)$ SCFT are realized, and matrix entanglement can be used to regularize the spatial entanglement in these QFTs. 
In Sec.~\ref{sec:BH-evaporation}, we consider the matrix entanglement for the evaporating small black hole. 
We show that the confined degrees of freedom play an important role. Sec.~\ref{sec:discussion} is devoted for conclusions and discussion of open problems. 

\section{Geometry encoded into matrix degrees of freedom}\label{sec:matrix-geometry}
In this section, we explain how the bulk geometry can be encoded into matrices, following Ref.~\cite{Hanada:2021ipb}. 
We consider the matrix model as a starter and then discuss the generalization to QFT. 

We consider the BFSS matrix model~\cite{Banks:1996vh} and simpler matrix models. One of the simplest is the bosonic part of the BFSS matrix model, 
\begin{align}
\hat{H}_{\rm BFSS, bos}
=
{\rm Tr}\left(
\frac{1}{2}\hat{P}_I^2
-
\frac{g^2}{4}[\hat{X}_I,\hat{X}_J]^2
\right). 
\end{align}
$I,J$ run from 1 to 9. The details of models do not matter unless we consider quantitative correspondence with gravity. 
Later, we will also consider the Gaussian matrix model as an even simpler model which can be solved exactly. 
Although the details of the Hamiltonian is not important, for concreteness let us consider the Hamiltonian interpolating the Gaussian matrix model and the bosonic part of the BFSS matrix model:
\begin{align}
\hat{H}
=
{\rm Tr}\left(
\frac{1}{2}\hat{P}_I^2
+
\frac{1}{2}\hat{X}_I^2
-
\frac{g^2}{4}[\hat{X}_I,\hat{X}_J]^2
\right). 
\label{eq:toy-Hamiltonian}
\end{align}
The Gaussian matrix model is obtained at $g^2=0$, and in the strong coupling limit $g^2\to\infty$ the mass term is negligible and the bosonic part of the BFSS matrix model is obtained. 
The mass term is analogous to the conformal mass term in 4d maximal super Yang-Mills compactified on three-sphere.

By using the generators $\tau_\alpha$ of U($N$) normalized as ${\rm Tr}(\tau_\alpha\tau_\beta)=\delta_{\alpha\beta}$, 
the matrix-valued operators are expressed as
\begin{align}
\hat{P}_{I,ij}
=
\sum_{\alpha=1}^{N^2}
\hat{P}_I^\alpha\tau_{\alpha,ij}, 
\qquad
\hat{X}_{I,ij}
=
\sum_{\alpha=1}^{N^2}
\hat{X}_I^\alpha\tau_{\alpha,ij}, 
\end{align}
with the canonical commutation relation
\begin{align}
[\hat{X}^\alpha_I,\hat{P}^\beta_J]
=
i\delta_{IJ}\delta^{\alpha\beta}. 
\end{align}
\subsection{Gauge-invariant Hilbert space and extended Hilbert space}\label{sec:H_inv-vs-H_ext}
To understand the matrix model in terms of quantum states, it is important to appreciate the difference of gauge-invariant Hilbert space ${\cal H}_{\rm inv}$ consisting only of gauge-singlet states and extended Hilbert space ${\cal H}_{\rm ext}$ that contains non-singlet states as well.
We can use the coordinate eigenstates $|X\rangle$ that satisfy $\hat{X}_{I,ij}|X\rangle=X_{I,ij}|X\rangle$ for all $9N^2$ combinations of $(I,i,j)$, as the basis of ${\cal H}_{\rm ext}$. 
Equivalently, we can use the momentum eigenstates $|P\rangle$ as the basis. 
Note that operators $\hat{X}_{I,ij}$ and $\hat{P}_{I,ij}$ are defined on this extended Hilbert space. 
Later in this paper, we will use wave-packet states as an over-complete basis of ${\cal H}_{\rm ext}$. 

The use of the extended Hilbert space is naturally justified if we use the projector from $\mathcal{H}_{\rm ext}$ to $\mathcal{H}_{\rm inv}$ that corresponds to the integration over Polyakov loop~\cite{Hanada:2020uvt,Hanada:2021ipb}. The partition function at finite temperature is written in two different but equivalent ways. The first expression is 
\begin{align}
Z(T)
&=
{\rm Tr}_{{\cal H}_{\rm inv}}\left(
e^{-\hat{H}/T}
\right). 
\label{eq:Z-H-inv-MM}
\end{align}
Based on this expression, people often say `physical states are gauge-singlets'. 
The second expression uses the extended Hilbert space that contains non-singlet modes:
\begin{align}
Z(T)
&=
\frac{1}{{\rm vol}G}\int_Gdg
{\rm Tr}_{{\cal H}_{\rm ext}}\left(
\hat{g}
e^{-\hat{H}/T}
\right). 
\label{eq:Z-H-ext-MM}
\end{align}
Here $G={\rm U}(N)$ is the gauge group, $g$ is a group element, and $\hat{g}$ is the representation of $g$ acting on the extended Hilbert space ${\cal H}_{\rm ext}$. The Haar measure is used for the integral.   
As shown in Appendix~\ref{sec:H-to-L}, the second expression is directly related to the path integral with temporal gauge field $A_t$, 
and $g\in G$ corresponds to the Polyakov loop. 
In this second expression, we can say that `states connected by a gauge transformation are identified', i.e., $|\phi\rangle$ and $\hat{g}|\phi\rangle$ are indistinguishable, just as we do in the classical theory or in the path-integral formulation. 

Note that 
\begin{align}
\hat{{\cal P}}
\equiv
\frac{1}{{\rm vol}G}\int_G dg
\hat{g}
\end{align}
is a projection operator from ${\cal H}_{\rm ext}$ to ${\cal H}_{\rm inv}$ that satisfies $\hat{{\cal P}}^2=\hat{{\cal P}}$. 
By using $\hat{{\cal P}}$, \eqref{eq:Z-H-inv-MM} can also be written as 
\begin{align}
Z(T)
&=
{\rm Tr}_{{\cal H}_{\rm ext}}\left(
\hat{{\cal P}}
e^{-\hat{H}/T}
\right). 
\end{align}
From a state $|\phi\rangle\in{\cal H}_{\rm ext}$, which can be non-singlet, we can obtain a singlet $\hat{{\cal P}}|\phi\rangle\in{\cal H}_{\rm inv}$. 
By using this correspondence, we can describe the same physics by using the singlet states or non-singlet states. 
Because $\hat{{\cal P}}$ and $\hat{H}$ commute with each other, this singlet-projection and Hamiltonian time evolution commute.  


\subsection{Geometry from wave packet in $\mathbb{R}^{9N^2}$}\label{sec:wave-packet}
\begin{figure}[htbp]
  \begin{center}
   \includegraphics[width=80mm]{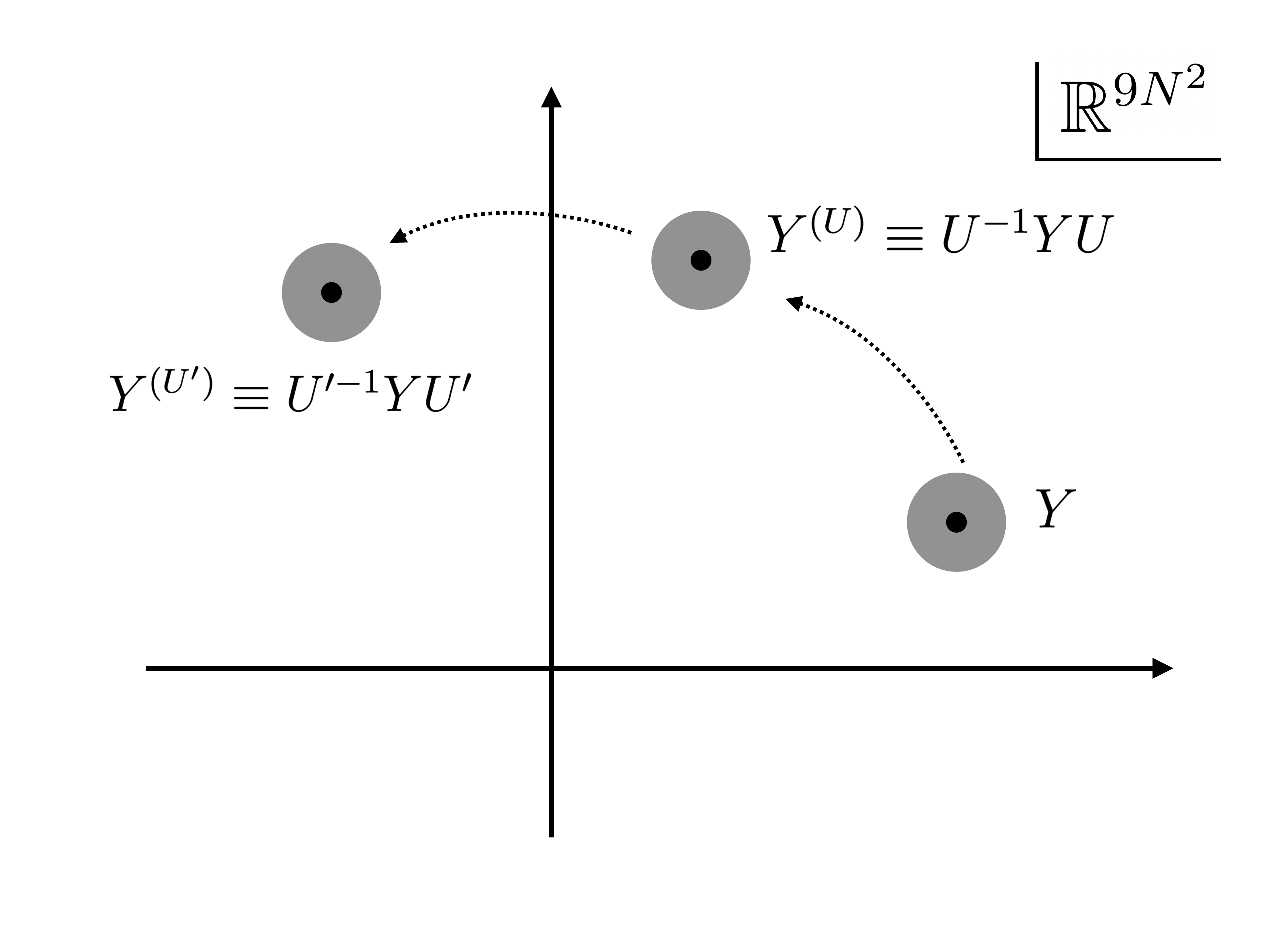}
  \end{center}
  \caption{
Wave packets in the coordinate space $\mathbb{R}^{9N^2}$. 
Black points are the centers of wave packets ($Y$, $Y^{(U)}$ and $Y^{(U')}$), and gray disks are the wave packets (more precisely, the regions where the wave functions are not vanishingly small). 
Under the gauge transformation, the center of the wave packet moves as $Y\to Y^{(U)}=U^{-1}YU$, but the shape of the wave packet does not change. The center of the wave packet gives `slow modes' that describe the geometry consisting of D-branes and strings. 
  }\label{fig:wave-packet}
\end{figure}

Low-energy states can be written as superpositions of wave packets. 
We consider a wave packet around $Y_I$ in coordinate space and $Q_I$ in momentum space.
Such a wave packet naturally describes geometry formed by D-branes and open strings~\cite{Hanada:2021ipb,Hanada:2021swb}, along the lines of Ref.~\cite{Witten:1995im}.  
A natural way to obtain a low-energy wave packet is to find $|Y,Q\rangle$ that satisfies constraints
\begin{align}
\langle Y,Q|\hat{X}_I|Y,Q\rangle = Y_I, 
\qquad
\langle Y,Q|\hat{P}_I|Y,Q\rangle = Q_I.
\end{align}
The state minimizing the energy
\begin{align}
\langle Y,Q|\hat{H}|Y,Q\rangle
\end{align}
is of particular interest. 
Note that such a $|Y,Q\rangle$ is an element of the extended Hilbert space ${\cal H}_{\rm ext}$, but not necessarily in the gauge-invariant Hilbert space ${\cal H}_{\rm inv}$. 
It transforms as 
\begin{align}
|Y,Q\rangle\to|U^{-1}YU,U^{-1}QU\rangle
\end{align}
via the gauge transformation; see Fig.~\ref{fig:wave-packet}. 
States that transform to each other via gauge transformation should be identified. 
Note that the shape of the wave packet does not change via gauge transformation. 
The only `matrices' that can be diagonalized are $Y_I$ and $Q_I$. 

\begin{figure}[htbp]
  \begin{center}
   \includegraphics[width=80mm]{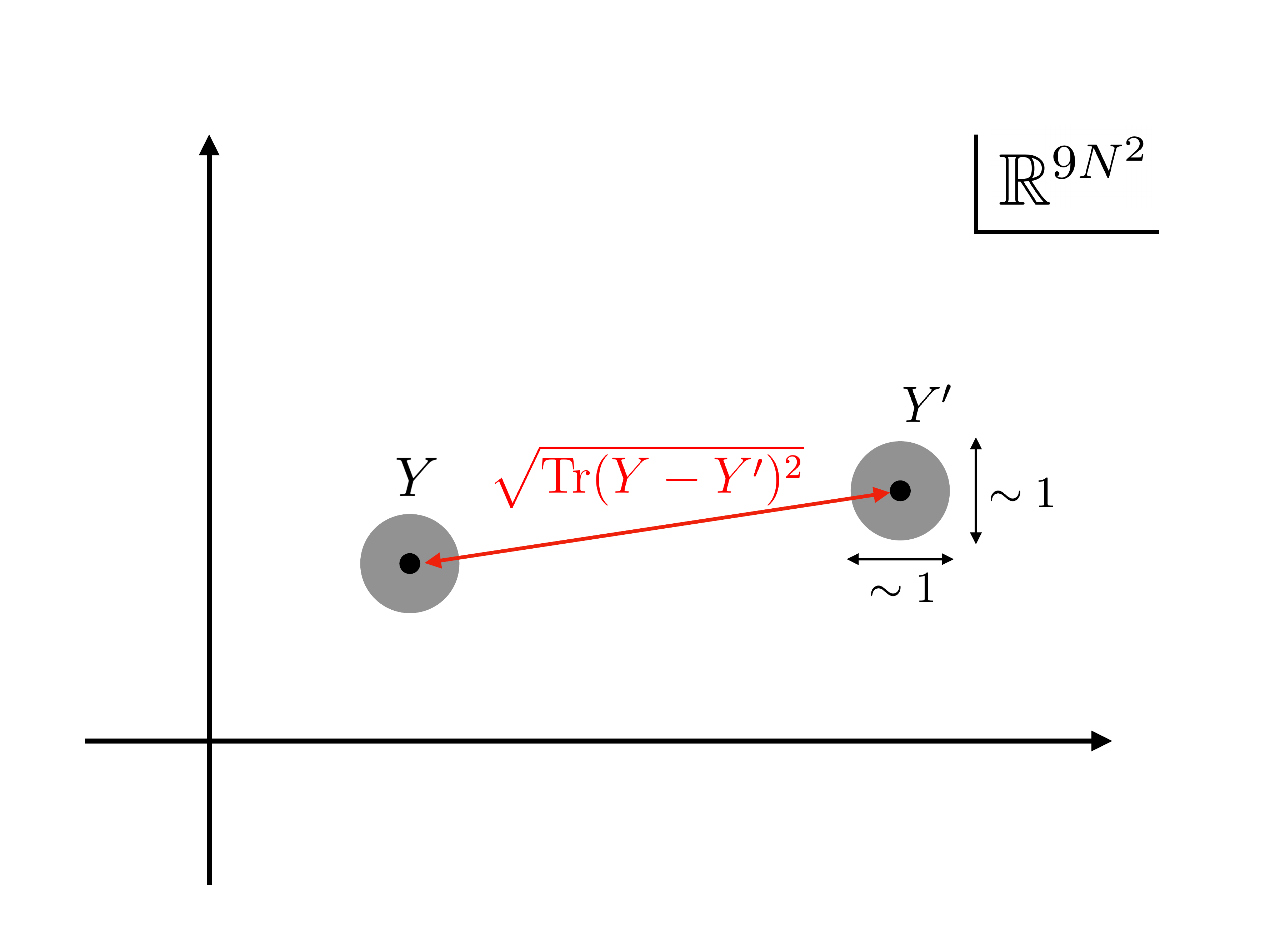}
  \end{center}
  \caption{
Each wave packet smoothly spreads in $\mathbb{R}^{9N^2}$. 
Along each of $9N^2$ dimensions, the size of the wave packet is at most of order $N^0$. 
Two wave packets centered around $Y$ and $Y'$ in $\mathbb{R}^{9N^2}$ have vanishingly small overlap if the centers of wave packets are well separated, i.e., $\sqrt{{\rm Tr}(Y-Y')^2}\gg 1$. 
  }\label{fig:wave-packet-distance}
\end{figure}

Each wave packet smoothly spreads in the coordinate space $\mathbb{R}^{9N^2}$, and also in the momentum space $\mathbb{R}^{9N^2}$.  
Along each of $9N^2$ dimensions, the size is at most of order $N^0$. 
Two wave packets centered around $Y$ and $Y'$ in $\mathbb{R}^{9N^2}$ have vanishingly small overlap if the centers of wave packets are well separated, i.e., $\sqrt{{\rm Tr}(Y-Y')^2}\gg 1$. See Fig.~\ref{fig:wave-packet-distance}. 

\subsubsection*{The simplest example: Gaussian matrix model and coherent state}
As an instructive example, let us consider the Gaussian matrix model that is obtained by setting the coupling in the toy model Hamiltonian~\eqref{eq:toy-Hamiltonian} to zero:
\begin{align}
\hat{H}
=
\sum_I
{\rm Tr}\left(
\frac{1}{2}\hat{P}_I^2
+
\frac{1}{2}\hat{X}_I^2
\right)
=
\frac{1}{2}
\sum_{I,\alpha}\left(\hat{P}_{I,\alpha}^2+\hat{X}_{I,\alpha}^2\right). 
\end{align}
Namely, we consider $9N^2$ harmonic oscillators subject to the U($N$) gauge symmetry. 
The ground state is simply the tensor product of $9N^2$ Fock vacua, 
\begin{align}
|0\rangle
\equiv
\otimes_{I,\alpha}|0\rangle_{I,\alpha}.
\end{align}
This ground state is U($N$)-invariant. 
The wave packet is defined above is the coherent state, 
\begin{align}
|Y,Q\rangle
=
e^{-i\sum_I{\rm Tr}(Y_I\hat{P}_I-Q_I\hat{X}_I)}|0\rangle. 
\end{align}
The ground state is a wave packet about $Y=0$ and $Q=0$, 
\begin{align}
|0\rangle
=
|Y=0,Q=0\rangle. 
\end{align}

Under the gauge transformation $\hat{X}_I,\hat{P}_I\to U\hat{X}_IU^{-1},U\hat{P}_IU^{-1}$, this wave packet transforms as  
\begin{align}
|Y,Q\rangle
\to
|Y^{(U)},Q^{(U)}\rangle
\equiv
|U^{-1}YU,U^{-1}QU\rangle.
\end{align}
The wave function is well localized around $Y_I$ in the coordinate space and $Q_I$ in the momentum space. For example, when $Q_I$ is zero, 
\begin{align}
\langle X|Y,Q=0\rangle
=
\frac{1}{\pi^{9N^2/4}}
\exp\left(
-\frac{1}{2}\sum_I{\rm Tr}(X_I-Y_I)^2
\right). 
\end{align}
The overlap between two wave packets is
\begin{align}
\langle Y',Q'=0|Y,Q=0\rangle
=
\exp\left(
-\frac{1}{8}\sum_I{\rm Tr}(Y_I-Y'_I)^2
\right). 
\end{align}
They have vanishingly small overlap if the centers of wave packets are well separated, i.e., $\sqrt{{\rm Tr}(Y-Y')^2}\gg 1$. 
In typical situations considered in holography, 
\begin{align}
1\ll ({\rm nonzero\ eigenvalues\ of}\ Y)\lesssim\sqrt{N}.  
\end{align}
Therefore, typical separation is much larger than 1, and overlap between typical wave packets is practically negligible. 
\subsubsection*{Geometric interpretation}
As explained in Refs.~\cite{Hanada:2021ipb,Hanada:2021swb}, matrices $Y_I$ are related to the geometry consisting of D-branes and open strings via Witten's interpretation~\cite{Witten:1995im}.\footnote{
The original proposal~\cite{Witten:1995im} concerns low-energy effective description of D-branes and open strings, and did not claim that such a interpretation is valid in gauge/gravity duality; actually Ref.~\cite{Witten:1995im} appeared before the duality was proposed. Matrix Theory proposal~\cite{Banks:1996vh} incorporated this viewpoint to the description of gravity. 
In AdS/CFT, the location of probe D-brane is described in this way~\cite{Maldacena:1997re}.
}$^,$\footnote{
In the past, people tried to apply Witten's interpretation to a generic coordinate eigenstate in a wave packet, and encountered pathological issues.
One has to consider a wave packet as a whole to extract the geometry.}
Let us list a few features justifying this:
\begin{itemize}
\item
Ref.~\cite{Witten:1995im} considered low-energy states, which should be wave packets. 
The only `matrices' associated with a wave packet that can be diagonalized are $Y_I$ and $Q_I$. 

\item
Suppose all 9 matrices can be diagonalized simultaneously: 
\begin{align}
(Y_I)_{ij}=y_{I,i}\delta_{ij},
\qquad
\vec{y}_i=(y_{1,i},\cdots,y_{9,i})\in\mathbb{R}^9. 
\end{align} 
Around this background, fluctuation of $(i,j)$ component has mass proportional to $|\vec{y}_i-\vec{y}_j|$.\footnote{More precisely, the transverse modes have mass, while the longitudinal modes corresponding to the motion of D-branes do not.
It is straightforward to check it analytically in the weak-coupling region of a toy model Hamiltonian~\eqref{eq:toy-Hamiltonian}.
}
This allows the interpretation that $\vec{y}_i$ is the location of the $i$-th D0-brane and off-diagonal entries describe the open strings stretched between D-branes.

\item
When the locations of D-branes coincide, the symmetry of the state is enhanced. 
If $N_1$ D-branes sit at $\vec{y}_1=\cdots=\vec{y}_{N_1}$, 
other $N_2$ D-barnes sit at $\vec{y}_{N_1+1}=\cdots=\vec{y}_{N_1+N_2}$, etc, the symmetry is ${\rm U}(N_1)\times{\rm U}(N_2)\times\cdots$. 
\item
If the matrices cannot be simultaneously diagonalized but can be taken to be simultaneously block diagonal, then each block describes an extended bound state of D-branes and open strings.
We can construct such states by choosing $Y$ and $Q$ appropriately. 

\end{itemize}

\subsection{Superselection at large $N$}
In the standard model of particle physics, ${\rm SU}(2)_{\rm L}\times{\rm U}(1)_{\rm Y}$ is `spontaneously broken' to $U(1)_{\rm EM}$ by the vacuum expectation value of the Higgs field. More precisely, the global part of gauge symmetry is broken~\cite{Elitzur:1975im} by a specific choice of boundary condition, although the local part is not.
Note that quantum states corresponding to different boundary condition are separated by superselection, because no gauge-invariant local operator can connect them. 
Formally, we can construct a gauge-invariant vacuum by taking a superposition of all possible boundary conditions.
However, due to the superselection, we cannot distinguish such a state from a gauge-symmetry-breaking vacuum with a specific boundary condition. 

Essentially the same mechanism for superselection applies to many states in large-$N$ gauge theories~\cite{Hanada:2019czd}. 
The wave packet $|Y,Q\rangle$ defined above is not U($N$)-invariant (unless $Y$ and $Q$ are proportional to the identity matrix). 
The gauge-invariant counterpart is obtained by symmetrizing over all possible U($N$) transformations, i.e., 
\begin{align}
\mathcal{N}^{-1/2}\int dU \hat{U}|Y,Q\rangle
=
\mathcal{N}^{-1/2}\int dU |U^{-1}YU,U^{-1}QU\rangle, 
\end{align}
where $\mathcal{N}$ is a normalization factor. 
In the large-$N$ limit, the size of the wave function along each direction $(I,\alpha)$ is of order $N^0$ or less, while typical eigenvalues of $Y_I$ and $Q_I$ we consider are of order $\sqrt{N}$. Even for a small U($N$) transformation (specifically, $g=e^{i\epsilon^\alpha\tau_\alpha}$, where $\sum_\alpha(\epsilon^\alpha)^2$ is small but of order $N^0$), the overlap between $|Y,Q\rangle$ and $|U^{-1}YU,U^{-1}QU\rangle$ is parametrically suppressed. 
In the free limit, the wave packets are coherent states, and we can explicitly see that the overlap behaves as $e^{-\frac{1}{8}\sum_I{\rm Tr}(Y_I-U^{-1}Y_IU)^2}$. 
In typical situations we are interested, ${\rm Tr}Y^2$ is sufficiently larger than $O(N^0)$. 
Therefore, the overlap is essentially the delta function. 
As observables, we consider gauge-invariant operators which does not change the energy of the system significantly, 
e.g., ${\rm Tr}(\hat{X}_{I_1}\hat{X}_{I_2}\cdots\hat{X}_{I_k})$, $k=O(N^0)$. Such operators cannot connect $|Y,Q=0\rangle$ and $|U^{-1}YU,Q=0\rangle$ unless $U$ is parametrically close to $G_Y$, where $G_Y\equiv\{g\in G| g^{-1}Yg=Y\}$ is the stabilizer of $Y$. 
Whether we take the expectation value of such an operator by using $|Y,Q\rangle$ or its gauge-invariant counterpart, we obtain the same value. 
In this sense, different gauge choice corresponds to different superselection sector. 
$|Y,Q=0\rangle$ and its gauge-symmetrization give the same expectation values, and hence, 
we cannot distinguish them. 

\subsection{Generalization to QFT}
Generalization to QFT (i.e., theories with nonzero spatial dimension) is straightforward.
The partition function is the same as \eqref{eq:Z-H-ext-MM}, except that we need to consider all possible local gauge transformation instead of $G$ that can be schematically written as as $\mathcal{G}=\otimes_{\vec{x}}G_{\vec{x}}$ where $G_{\vec{x}}$ is the gauge group $G$ acting on a spatial point $\vec{x}$. 
Thermal partition function can be written as 
\begin{align}
Z(T)
&=
\frac{1}{{\rm vol}\mathcal{G}}\int_\mathcal{G}dg
{\rm Tr}_{{\cal H}_{\rm ext}}\left(
\hat{g}
e^{-\hat{H}/T}
\right). 
\label{eq:Z-H-ext-QFT}
\end{align}

As a concrete example, let us consider $(p+1)$-dimensional maximal super Yang-Mills theory. 
If we use a lattice regularization for the spatial dimensions, wave functions are defined on $\mathbb{R}^{(9-p)\times N^2\times n_{\rm site}}$, where $9-p$ is the number of scalar fields and $n_{\rm site}$ is the number of spatial lattice sites. 
(Strictly speaking, we have to take into account the spatial directions of the gauge field, whose discretized version live on the links.) The center of a wave packet is specified by $Y_I(\vec{x})$ and $Q_I(\vec{x})$. 

In the case of $p=3$, D-brane geometry on $\mathbb{R}^6=\mathbb{R}_{>0}\times {\rm S}^5$ can be generated from six scalar fields. $\mathbb{R}_{>0}$ is interpreted as the radial coordinate of AdS$_5$. 
In the ground state, $Y_I=0$ and $Q_I=0$, and hence, all D-branes are sitting at the origin of AdS$_5$, and no open string is excited. 
This is exactly the expected property of the BPS limit of black three-brane. 

\subsection{Examples}

Let us list a few examples of the encoding of geometry into matrix degrees of freedom. 
We consider matrix model for simplicity. Generalization to QFT is straightforward.  

\subsubsection{Probe D-branes in BPS black hole geometry}
The ground state is given by $Y=0$, $Q=0$.~\footnote{
In the case of BFSS matrix model~\cite{Banks:1996vh}, $Y_I=c_I\cdot\textbf{1}$, 
where $\textbf{1}$ is the $N\times N$ unit vector and $c_I$ is any real number, is equally fine. If we impose the traceless condition, $c_I$ has to be zero.
} 
All D-branes are sitting on top of each other at the origin, no string is excited, and the area of the horizon is zero. 

We can separate a few D-branes from others and probe the geometry. 
Let us take the BFSS matrix model as an example, and take the $(N,N)$ components $\vec{y}\equiv(Y_{1,NN},\cdots,Y_{9,NN})$, $\vec{q}\equiv(Q_{1,NN},\cdots,Q_{9,NN})$ nonzero. 
Such wave packet describes the BPS black zero-brane and a probe D-brane. 
Typical distance scale considered for the duality to type IIA black zero-brane is small but of order $\sqrt{N}$ (i.e., $\epsilon\sqrt{N}$, where $\epsilon$ is small but of order $N^0$). 
When $|\vec{y}|$ is of this order, $(1+9)$-dimensional geometry is probed. 
As $|\vec{y}|$ becomes smaller, string coupling becomes larger, and eventually $(1+10)$-dimensional geometry sets in~\cite{Itzhaki:1998dd}.  
If we also take the $(N-1,N-1)$ components nonzero, the scattering of two D-branes in the black zero-brane geometry can be described.  
\subsubsection{Two-black-hole geometry and thermofield double state}\label{sec:2-BH-TFD}
Let us consider states corresponding to two stacks of D-branes separated by a distance $L$.
We write that the center of wave packet $Y_I$ and $Q_I$ to be block diagonal,
\begin{align}
Y_I
=
\left(
\begin{array}{cc}
Y_I^{(1)} & 0 \\
0 & Y_I^{(2)}
\end{array}
\right), 
\qquad
Q_I
=
\left(
\begin{array}{cc}
Q_I^{(1)} & 0 \\
0 & Q_I^{(2)}
\end{array}
\right), 
\label{block-diag}
\end{align}
where $Y^{(1)}$ and $Q^{(1)}$ are $N_1\times N_1$, and $Y^{(2)}$ and $Q^{(2)}$ are $N_2\times N_2$. Furthermore, we assume
\begin{align}
& 
Y^{(1)}_1=\frac{L}{2}\cdot\textbf{1}+Y_1^{\prime(1)}, 
\qquad
Y^{(2)}_1=-\frac{L}{2}\cdot\textbf{1}+Y_1^{\prime(2)}, 
\nonumber\\
&
Y^{(1)}_{I=2,\cdots,9}=Y_{I=2,\cdots,9}^{\prime(1)}, 
\qquad
Y^{(2)}_{I=2,\cdots,9}=Y_{I=2,\cdots,9}^{\prime(2)}, 
\nonumber\\
&
Q^{(1)}_{I=1,\cdots,9}=Q_{I=1,\cdots,9}^{\prime(1)}, 
\qquad
Q^{(2)}_{I=1,\cdots,9}=Q_{I=1,\cdots,9}^{\prime(2)}, 
\end{align}
where the eigenvalues of $Y^\prime$ and $Q^\prime$ are negligibly small compared to $L$. 
  
Let us use the following notation:
\begin{align}
\hat{X}_1
=
\left(
\begin{array}{cc}
\hat{X}_1^{\prime(1)} + \frac{L}{2}\cdot\textbf{1} & \hat{Z}^\prime \\
\hat{Z}^{\prime\dagger} & \hat{X}_1^{\prime(2)} - \frac{L}{2}\cdot\textbf{1}
\end{array}
\right), 
\qquad
\hat{X}_{I=2,\cdots,9}
=
\left(
\begin{array}{cc}
\hat{X}_I^{\prime(1)} & \hat{W}^\prime_I \\
\hat{W}_I^{\prime\dagger} & \hat{X}_I^{\prime(2)}
\end{array}
\right), 
\end{align}
\begin{align}
\hat{P}_{I=1,\cdots,9}
=
\left(
\begin{array}{cc}
\hat{P}_I^{\prime(1)} & \hat{R}^\prime_{I} \\
\hat{R}_{I}^{\prime\dagger} & \hat{P}_I^{\prime(2)}
\end{array}
\right).  
\end{align}
If $L$ is sufficiently large, $\hat{W}^\prime$s acquire mass of order $L$ and decouple. More precisely, these off-diagonal blocks, together with $\hat{R}^\prime_{I=2,\cdots,9}$, can be treated as harmonic oscillators with large frequency of order $L$, and they get frozen to the ground state. $\hat{Z}^\prime$ does not acquire mass, but it enters the Hamiltonian only via the interaction with $\hat{W}_{I=2,\cdots,9}$, and hence, it also decouples. 
Then we have copies of two matrix models with SU($N_1$) or SU($N_2$) gauge group. 
(Note that, when off-diagonal blocks decouple, $L$ is so large that dual gravity description in terms of black zero brane geometry is not valid.)
This setup is analogous to the ones studied in Refs.~\cite{Mollabashi:2014qfa,Karch:2014pma}.
If diagonal blocks have sufficiently large energy, each of them can be regarded as a black hole. 
\subsubsection*{Thermofield double}

If the two blocks are infinitely separated, off-diagonal entries completely decouple from the dynamics. 
Then we have two decoupled Hilbert spaces. 
The Hamiltonian turns to the sum of the SU($N_1$) Hamiltonian and SU($N_2$) Hamiltonian. 
When $N_1=N_2$, this is exactly the setup of two copies of same Hilbert spaces with the same Hamiltonian, considered in Ref.~\cite{Maldacena:2001kr}. 
By taking certain linear combinations of such two-stack wave-packet states, the thermofield double states can be constructed. 

\subsubsection{Small black hole (partially-deconfined phase)}\label{sec:partially-deconfined-state}
A small black hole can be described by using partially-deconfined states~\cite{Hanada:2016pwv,Berenstein:2018lrm,Hanada:2018zxn,Hanada:2019czd,Hanada:2020uvt}, in which a subgroup of U($N$) is deconfined. 
Roughly speaking, the SU($N_1$)-deconfined states can be constructed as linear combinations of wave packets of the form \eqref{block-diag}, with $Y^{(2)}=Q^{(2)}=0$. 
Namely, we consider 
\begin{align}
Y_I
=
\left(
\begin{array}{cc}
 Y^{(1)}_I& 0\\
0& 0
\end{array}
\right), 
\qquad
Q_I
=
\left(
\begin{array}{cc}
 Q^{(1)}_I& 0\\
0& 0
\end{array}
\right). 
\label{partially-deconfined-state}
\end{align}
We assume $Y^{(1)}_I$ and $Q^{(1)}_I$ are significantly excited, i.e., ${\rm Tr}(Y^{(1)}_I)^2\sim {\rm Tr}(Q^{(1)}_I)^2\sim N_1^2$, while other blocks are set to zero.
Both $N_1$ D-branes and open strings between them are excited and form an extended object. 
Other $N_2=N-N_1$ D-branes are sitting at the origin and no open strings are excited between them, or between those $N_2$ D-branes and the other $N_1$ D-branes. Note that all D-branes contribute to the exterior geometry. 
Such states can be interpreted as a small black hole, with no radiation emitted yet. 

To describe the Hawking radiation, we can take $Y^{(2)}$ and $Q^{(2)}$ to be nonzero, and add small blocks whose sizes are of order one, and interpret them as gravitons. The idea behind it is that the large and small blocks are long and short strings, which are black hole and graviton, respectively.
To understand it in terms of operators, let us consider $\hat{O}_{IJ}\equiv{\rm Tr}(\hat{X}_{I}\hat{X}_{J})$ as an example. 
By definition, $\hat{O}_{IJ}=\sum_{i,j}\hat{X}_{I}^{ij}\hat{X}_{J}^{ji}$. 
Terms with $i=j$ is negligible at large $N$, and hence, 
$\hat{O}_{IJ}=\hat{X}_{I}^{12}\hat{X}_{J}^{21}+{\rm (permutations)}$. Therefore, with an appropriate gauge fixing, such an operator excites a $2\times 2$ block which can be seen as a closed string made of two open strings. In the same manner, from a trace of product of $L$ operators, we can get a closed string made of $L$ open strings.
In this way, the block-diagonal structure naturally appears after gauge fixing. 
Note that $N_1$ decreases as the black hole shrinks, and the number of small blocks (number of gravitons emitted from black hole) grows. 
\subsubsection{Fuzzy sphere}\label{sec:fuzzy_sphere_definition}
Another important class of bound states of D-branes is fuzzy sphere. Fuzzy sphere appears through the Myers effect~\cite{Myers:1999ps} when D-branes are coupled to background flux. As a concrete setup, here we consider fuzzy sphere in the BMN matrix model~\cite{Berenstein:2002jq}. In this case, fuzzy spheres can describe D2-branes, M2-branes, NS5-branes or M5-branes, depending on the way the large-$N$ limit is taken~\cite{Maldacena:2002rb,Asano:2017nxw,Asano:2017xiy}. Intuitively, fuzzy spheres are analogous to a lattice regularization. Spatial geometry of such higher-dimensional objects are encoded into matrix degrees of freedom, and hence, the matrix entanglement in the BMN matrix model can be reinterpreted as the spatial entanglement in the world-volume theory on the higher-dimensional objects; see Sec.~\ref{sec:fuzzy_sphere}.  

The bosonic part of the BMN matrix model is written as
\begin{align}
\hat{H}_{\rm BMN, bos}
=
{\rm Tr}\left(
\frac{1}{2}\hat{P}_I^2
+
\frac{1}{18}\hat{X}_i^2
+
\frac{1}{72}\hat{X}_a^2
+
\frac{\sqrt{-1}\mu g}{3}\epsilon^{ijk}\hat{X}_i\hat{X}_j\hat{X}_k
-
\frac{g^2}{4}[\hat{X}_I,\hat{X}_J]^2
\right),  
\end{align}
where $I,J=1,2,\cdots,9$,  $i,j,k=1,2,3$, $a=4,5,\cdots,9$.
$\epsilon^{ijk}$ is the structure constant of SU(2) normalized as $\epsilon^{123}=1$. 
The fuzzy-sphere background is a wave packet localized about the SU(2) algebra, 
\begin{align}
Y_i = \frac{\mu}{3g}J_{i}, 
\end{align}
where
\begin{align}
[J_i,J_j]=\sqrt{-1}\epsilon^{ijk}J_k.  
\end{align}
There are various representations of SU(2), and hence there are many fuzzy sphere backgrounds that are BPS ground states. Let us take $J_i$ to be $N_2$ copies of $N_5$-dimensional irreducible representations, 
\begin{align}
J_i
=
J_i^{(N_5)}\otimes\textbf{1}_{N_2}, 
\end{align}
where $N=N_2N_5$. 
When $N_2$ is fixed and $N_5$ is sent to infinity, the system of $N_2$-coincident D2- or M2-branes are described. (Coupling constant $g$ and flux parameter $\mu$ should be scaled appropriately with $N$.) On the other hand, when $N_5$ is fixed and $N_2$ is sent to infinity, the system of $N_5$-coincident NS5- or M5-branes are described.

\section{Entanglement between matrix degrees of freedom (Matrix Entanglement)}\label{sec:EE-definition}
In this section, we define the matrix entanglement. 
We utilize the geometric interpretation reviewed in Sec.~\ref{sec:matrix-geometry}. 
Although we consider only `matrix' entanglement here, the generalization to theories with generic matter content, e.g., theories with a field in the fundamental representation such as QCD, is straightforward.   
\subsection{Matrix model}
Let us start with a rather formal argument. 
Suppose $N^2$ generators of U($N$) are split into two groups $A=\{\alpha\}$ and $\bar{A}=\{\alpha'\}$. The extended Hilbert space ${\cal H}_{\rm ext}$ factorizes as 
\begin{align}
{\cal H}_{\rm ext}={\cal H}_A\otimes{\cal H}_{\bar{A}}, 
\end{align}
where 
\begin{align}
{\cal H}_A
\equiv
{\rm Span}\{\otimes_{\alpha}|X_\alpha\rangle | \alpha\in A, X_\alpha\in\mathbb{R}\}, 
\qquad
{\cal H}_{\bar{A}}
\equiv
{\rm Span}\{\otimes_{\alpha'}|X_{\alpha'}\rangle| \alpha'\in\bar{A}, X_{\alpha'}\in\mathbb{R}\}.  
\end{align}
Therefore, by tracing out ${\cal H}_A$ or ${\cal H}_{\bar{A}}$, we can define the entanglement entropy. 

A nontrivial issue is if there is a natural separation of colors into $A$ and $\bar{A}$. 
The answer is positive, as we have seen in Sec.~\ref{sec:matrix-geometry}. Namely, there are states that admit geometric picture relating the matrix eigenvalues and bulk coordinate, based on the wave-packet states.  
One might be worried about an apparent need for the gauge fixing. 
However, such a `gauge fixing' is automatic at large $N$ thanks to the `spontaneous breaking' of U($N$), or more precisely the emergence of superselection sectors.

When the superselection takes place, the situation is analogous to the system of identical particles which are well separated. Let us consider the system of two indistinguishable bosons with wave function
\begin{align}
\Psi(\vec{x}_1,\vec{x}_2)
=
\frac{\phi(\vec{x}_1)\chi(\vec{x}_2)+\phi(\vec{x}_2)\chi(\vec{x}_1)}{\sqrt{2}}, 
\end{align} 
where the one-particle wave functions $\phi$ and $\chi$ have vanishingly small overlap due to the separation in space or energy barrier. In such a case, symmetrization of the wave function does not have physical effects, and apparent `entanglement' coming from the symmetrization of a product state should not have physical effect either. 
(See e.g.,~\cite{Schliemann2001} regarding this point.) 
By replacing the permutation group with U($N$) and repeating the same argument, we should not allow apparent `entanglement' due to the symmetrization over U($N$). 
A natural procedure is to take a superselection sector and define the entanglement in that sector. 

As an example, let us consider two stacks of D-branes described by a wave packet in which $Y_I$'s and $Q_I$'s are block-diagonal as in \eqref{block-diag}. 
Off-diagonal blocks mediate interaction between diagonal blocks. 
Whether the off-diagonal blocks are treated separately or not is a nontrivial issue. 
One option is to combine them with one of the diagonal blocks. 
Then the factorization of the Hilbert space is schematically expressed as 
\begin{align}
\left(
\begin{array}{cc}
\mathcal{H}_A & \mathcal{H}_{\bar{A}}\\
 \mathcal{H}_{\bar{A}} & \mathcal{H}_{\bar{A}}
\end{array}
\right). 
\label{matrix-partitioning-A-Abar}
\end{align}
For this, we can define the entanglement entropy as
\begin{align}
S_A
=
-{\rm Tr}_A\left(
\hat{\rho}_A\log\hat{\rho}_A
\right), 
\qquad
\hat{\rho}_A
=
{\rm Tr}_{\bar{A}}\hat{\rho},  
\label{MM-entanglement-def-1}
\end{align}
where $\hat{\rho}$ is the density matrix for the entire system. 
Equivalently, this is the von Neumann entropy of region $A$. 
If the density matrix describing the entire system is pure, then $S_A=S_{\bar{A}}$. 

Another option is to treat the off-diagonal blocks separately, 
\begin{align}
\left(
\begin{array}{cc}
\mathcal{H}_A & \mathcal{H}_C\\
 \mathcal{H}_C & \mathcal{H}_B
\end{array}
\right).
\label{matrix-partitioning-A-B-C}
\end{align}
Then, we can first trace out $\mathcal{H}_C$ to obtain the reduced density matrix for diagonal blocks as
\begin{align}
\hat{\rho}_{A\cup B}
=
{\rm Tr}_C\hat{\rho}.    
\end{align}
The reduced density matrix $\hat{\rho}_{A\cup B}$ is not necessarily a pure state, i.e., 
\begin{align}
  S_{A\cup B}=-{\rm Tr}_{A\cup B}\hat{\rho}_{A\cup B}\log\hat{\rho}_{A\cup B}
\end{align} 
can be nonzero. 
A natural measure for the entanglement between region $A$ and region $B$ would be the mutual information $S_A+S_B-S_{A\cup B}$, where 
  \begin{align}
  & S_A=-{\rm Tr}_A\hat{\rho}_A\log\hat{\rho}_A,
  \qquad 
  \hat{\rho}_A
  =
  {\rm Tr}_B\hat{\rho}_{A\cup B}
  = 
  {\rm Tr}_{B\cup C}\hat{\rho},
  \nonumber\\
  & S_B = -{\rm Tr}_B\hat{\rho}_B\log\hat{\rho}_B, 
  \qquad
  \hat{\rho}_B
  =
  {\rm Tr}_A\hat{\rho}_{A\cup B}
  =
  {\rm Tr}_{A\cup C}\hat{\rho}. 
  \end{align}

When the theory under consideration has weakly-coupled gravity dual and the partitioning of matrix degrees of freedom corresponds to geometric partitioning on the gravity side, it would be natural to expect that the matrix entanglement is calculated by using holographic fine-grained entropy formulas~\cite{Ryu:2006bv,Hubeny:2007xt,Lewkowycz:2013nqa,Faulkner:2013ana,Engelhardt:2014gca}.  
Depending on the geometry encoded into matrices, there can be multiple ways that matrix degrees of freedom are split into pieces. Then, different partitioning of matrix degrees of freedom should correspond to different extremal surfaces on the gravity side. 
In Sec.~\ref{sec:BH-evaporation}, we will consider black hole and Hawking radiations as such an example. 

Note that we do not find (or at least have not found yet) a natural way of splitting colors into two sectors 
when a given state is gauge-invariant even in the extended Hilbert space, such as the confining vacuum that corresponds to $Y=Q=0$. 
This is very different from the target-space entanglement whose main motivation includes the application to the ground state~\cite{Das:2020jhy,Das:2020xoa,Hampapura:2020hfg};  see Sec.~\ref{sec:target-space-EE}.

\subsection{Generalization to QFT}\label{sec:QFT}
In the past, the definition of the spatial entanglement in lattice gauge theory has been considered. Refs.~\cite{Ghosh:2015iwa,Soni:2015yga,Aoki:2015bsa} introduced the extended Hilbert space ${\cal H}_{\rm ext}$ because otherwise the factorization of the Hilbert space does not hold, that is, 
\begin{align}
{\cal H}_{{\rm ext},A\cup\bar{A}}
=
{\cal H}_{{\rm ext},A}
\otimes
{\cal H}_{{\rm ext},\bar{A}}
\label{eq:factorization-Hilbert-space}
\end{align}
while 
\begin{align}
{\cal H}_{{\rm inv},A\cup\bar{A}}
\supsetneq
{\cal H}_{{\rm inv},A}
\otimes
{\cal H}_{{\rm inv},\bar{A}}, 
\end{align}
where $A$ is a spatial region and $\bar{A}$ is its complement. 

We use the same extended Hilbert space ${\cal H}_{\rm ext}$ to define the matrix entanglement. The use of ${\cal H}_{\rm ext}$ is crucial not just for the factorization of the Hilbert space, but also for the geometric interpretation, as explained in Sec.~\ref{sec:matrix-geometry}. 

With a physically-well-motivated gauge fixing, we can define the entanglement entropy by splitting the matrix degrees of freedom (the internal space) into two pieces. In general, different partitionings of matrix entries are allowed at different spatial regions. 
The factorization \eqref{eq:factorization-Hilbert-space} holds for such generic partitions as well. 
In this paper, we consider only the partitioning of color space into two pieces. 

\subsection{Target-space entanglement vs matrix entanglement}\label{sec:target-space-EE}

To understand the difference between matrix entanglement and target-space entanglement, and also some similarities in special situations, let us briefly review target-space entanglement.

Suppose there are $N$ particles in $\mathbb{R}^3$. We fix $N$ and work in the first-quantized setup. There are $3N$ coordinate operators $\hat{x}_i, \hat{y}_i$ and $\hat{z}_i$ ($i=1,2,\cdots,N$), where a set of three operators $(\hat{x}_i,\hat{y}_i,\hat{z}_i)$ defines the location of $i$-th particle in `target space' $\mathbb{R}^3$. When $N$ particles are indistinguishable, the states transforming to each other by S$_N$ permutation are identified. 
Namely, the S$_N$-permutation
\begin{align}
(\hat{x}_i,\hat{y}_i,\hat{z}_i)\to(\hat{x}_{\sigma(i)},\hat{y}_{\sigma(i)},\hat{z}_{\sigma(i)}), 
\qquad 
\sigma\in{\rm S}_N
\end{align}
is a gauge symmetry. 
Up to the gauge redundancy, the $N$-particle Hilbert space can be written as
\begin{align}
\mathcal{H}^{(N)}
=
{\rm Span}
\left\{
|x_1,y_1,z_1;x_2,y_2,z_2;\cdots;x_N,y_N,z_N\rangle|
(x_i,y_i,z_i)\in\mathbb{R}^3
\right\}. 
\end{align}
We can split the target space $\mathbb{R}^3$ into two pieces $A$ and $\bar{A}$, 
and consider a subspace $\mathcal{H}^{(N)}_m\subset\mathcal{H}^{(N)}$ consisting of states in which $m$ particles are in $A$ and $N-m$ particles are in $\bar{A}$. 
Written explicitly, 
\begin{align}
\mathcal{H}_m^{(N)}
=
\mathcal{H}_{A,m}
\otimes
\mathcal{H}_{\bar{A},N-m},   
\end{align}
where
\begin{align}
\mathcal{H}_{A,m}
=
{\rm Span}
\left\{
|x_1,y_1,z_1;\cdots;x_m,y_m,z_m\rangle|
(x_i,y_i,z_i)\in A\subset\mathbb{R}^3
\right\}
\end{align}
and
\begin{align}
\mathcal{H}_{\bar{A},N-m}
=
{\rm Span}
\left\{
|x_{m+1},y_{m+1},z_{m+1};\cdots;x_N,y_N,z_N\rangle|
(x_i,y_i,z_i)\in\bar{A}\subset\mathbb{R}^3
\right\}. 
\end{align} 
Note that the gauge symmetry is fixed down to S$_m\times$S$_{N-m}$ in each $\mathcal{H}_m$. 
The full Hilbert space $\mathcal{H}^{(N)}$ can be written as
\begin{align}
\mathcal{H}^{(N)}
=
\oplus_{m=0}^N
\mathcal{H}_m^{(N)}
\equiv
\oplus_{m=0}^N
\left(
\mathcal{H}_{A,m}
\otimes
\mathcal{H}_{\bar{A},N-m}
\right).  
\end{align}
For a density matrix $\hat{\rho}$ defined on $\mathcal{H}^{(N)}$, we can trace out $\mathcal{H}_{\bar{A},N-m}$ for each $\mathcal{H}_m^{(N)}$, to define a reduced density matrix $\hat{\rho}_A$ defined on $\oplus_{m=0}^N\mathcal{H}_{A,m}$. 
Equivalently, we can regard $\hat{\rho}$ as an operator acting on the second-quantized Hilbert space $\sum_{N=0}^\infty\mathcal{H}^{(N)}$, which satisfies a factorization
\begin{align}
\sum_{N=0}^\infty\mathcal{H}^{(N)}
=
\mathcal{H}_{A}\otimes \mathcal{H}_{\bar{A}},
\end{align}
where $\mathcal{H}_A\equiv\oplus_{m=0}^\infty\mathcal{H}_{A,m}$ and $\mathcal{H}_{\bar{A}}\equiv\oplus_{m=0}^\infty\mathcal{H}_{\bar{A},m}$.
Then we can obtain $\hat{\rho}_A$ by tracing out $\mathcal{H}_{\bar{A}}$. 
The von Neumann entropy defined by using this $\hat{\rho}_A$ is the `target-space entanglement entropy'~\cite{Mazenc:2019ety}. It is possible to define such quantity in a gauge-invariant manner. 

Such a definition is motivated by the expectation that partitioning of target space captures important physics. This is often true, and hence the target space entanglement can provide us with useful insight. 
However, the partitioning of target space is not always the right way to go. For example, superfluid and normal fluid are not spatially separated. Still, if there is a natural way to separate the system into two pieces such as some sort of superselection, then entanglement between these two sectors can be defined. In the `particle entanglement' approach~\cite{wiseman2003entanglement}, superselection associated with particle number is used. 

Next, let us consider U($N$)-gauged matrix models. For one-matrix model~\cite{Das:1995vj,Hartnoll:2015fca,Sugishita:2021vih}, a natural approach is to fix U($N$) by diagonalizing the matrix. Then the system is interpreted as a system of $N$ particles with the residual S$_N$ symmetry, and hence, we can define target-space entanglement just as above. Things get complicated when there are two or more matrices, say $X_1,\cdots,X_d $ where $d\ge 2$. In the past, it has been proposed that one should diagonalize one of the matrices, say $X_1$, and use its eigenvalues to specify two regions $A$ and $\bar{A}$ on $\mathbb{R}$~\cite{Das:2020jhy,Das:2020xoa,Hampapura:2020hfg}. 
Although such a definition is well-defined, it is not clear if the entanglement entropy defined in this way correctly captures emergent geometry in the context of holography. In fact, at least for low-energy states including the ground state, the approach based on wave packets captures the geometry of string theory motivated by the probe-D-brane picture more precisely, as we have seen in Sec.~\ref{sec:matrix-geometry}.

\section{Application 1: Noncommutative geometry, M2-branes and M5-branes}\label{sec:fuzzy_sphere}
As the first example of application of matrix entanglement, we consider fuzzy sphere in the BMN matrix mnodel, which was introduced in Sec.~\ref{sec:fuzzy_sphere_definition}. We consider small excitations above the fuzzy-sphere background $Y_i = \frac{\mu}{3g}J_{i}$. We take $J_i$ to be $N_2$ copies of $N_5$-dimensional irreducible representation, $J_i=J_i^{(N_5)}\otimes\textbf{1}_{N_2}$. Then, in the appropriate large-$N$ limit, D2/M2-brane theory (3d maximal SYM) or NS5/M5-brane theory (6d $\mathcal{N}=(2,0)$ theory) can be described. 
The BMN matrix model can be used to regularize these theories, similarly to the lattice gauge theory. As we will see shortly, geometric information is encoded into matrix degrees of freedom. Therefore, by using matrix entanglement, we can define the regularized version of spatial entanglement.

Firstly, let us consider the large-$N$ limit that describes $N_2$-coincident D2- or M2-branes, i.e., fixed $N_2$ and $N_5\to\infty$. In this limit, 3d maximal SYM on two-sphere is obtained. More precisely, the radius of sphere is $\frac{3}{\mu}$, the coupling constant is $g_{\rm 3d}^2=\frac{g_{1d}^2}{\mu^2N_5}$, and spatial coordinate has noncommutativity characterized by the noncommutativity parameter $\theta=\frac{1}{\mu^2N_5}$. If the noncommutativity parameter is sent to zero, the ordinary 3d SYM on commutative sphere is obtained. (The commutative limit is not singular because of maximal supersymmetry~\cite{Matusis:2000jf,Hanada:2014ima}.)

Geometric information on two-sphere is encoded into matrix degrees of freedom via the fuzzy spherical harmonics; see e.g.,~Ref.~\cite{Iso:2001mg}. Therefore, geometric partitioning on two-sphere can be mapped into a partitioning of matrix degrees of freedom. Refs.~\cite{Karczmarek:2013jca,Okuno:2015kuc,Chen:2017kfj} applied this idea to scalar theories on fuzzy sphere. They considered one-matrix model whose kinetic term is proportional to ${\rm Tr}(\dot{\Phi}^2+\frac{\mu^2}{9}\sum_{i=1}^3[J_i,\Phi]^2+m^2\Phi^2)$, where $\Phi$ is an $N\times N$ Hermitian matrix. Note that $J_i$'s are introduced by hand. 
Our matrix-entanglement approach generalize it to gauge theory, in which $J_i$ emerges dynamically.  

\begin{figure}[htbp]
  \begin{center}
   \includegraphics[width=80mm]{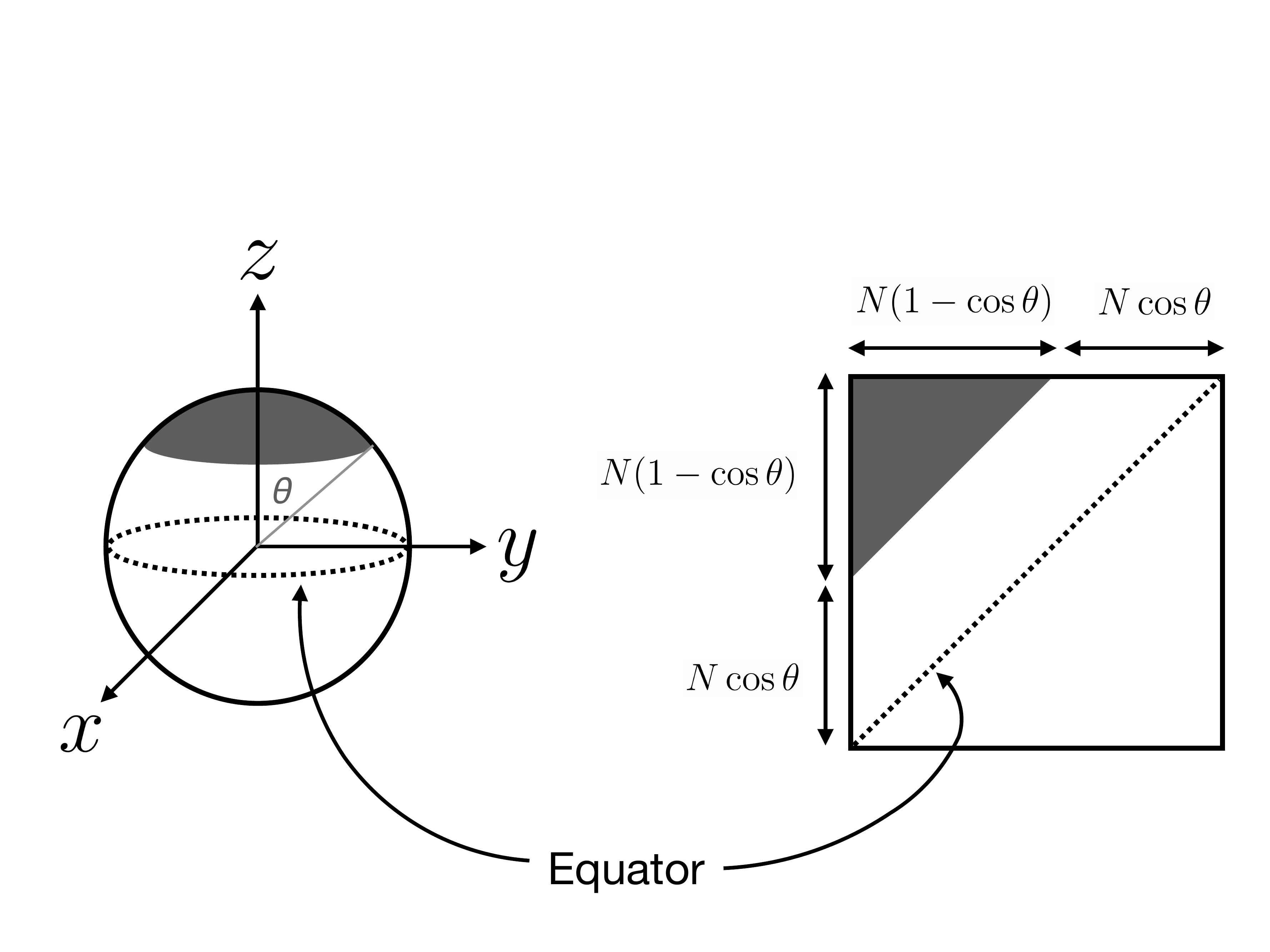}
  \end{center}
  \caption{
Partitioning of fuzzy sphere based on latitude (left) and the corresponding partitioning of matrix degrees of freedom (right)~\cite{Karczmarek:2013jca,Okuno:2015kuc,Chen:2017kfj}. 
Specific gauge choice \eqref{SU(2)-gauge-fixing} is assumed. 
  }\label{fig:Fuzzy_sphere}
\end{figure}

Specifically, let us fix the gauge such that $J_3$ is diagonalized in such a way that $J_3^{(N_5)}={\rm diag}(s,s-1,\cdots,-s+1,-s)$ and
\begin{align}
J_3
=
\left(
\begin{array}{ccccc}
s \cdot\textbf{1}_{N_2}& & & & \\
& (s-1)\cdot\textbf{1}_{N_2} & & & \\
& &  \ddots & & \\
& & & (-s+1)\cdot\textbf{1}_{N_2}& \\
& & & &-s\cdot\textbf{1}_{N_2}
\end{array}
\right), 
\label{SU(2)-gauge-fixing}
\end{align}
where $2s+1=N_5$.
Then, if we split fuzzy sphere into two spatial regions as shown in the left panel in Fig.~\ref{fig:Fuzzy_sphere}, the corresponding partitioning of matrix degrees of freedom becomes like the right panel of Fig.~\ref{fig:Fuzzy_sphere}~\cite{Karczmarek:2013jca,Okuno:2015kuc,Chen:2017kfj}. 
Note that there is no ambiguity in the partitioning of the matrix side, including the off-diagonal entries. 

Next, let us consider another large-$N$ limit that describes $N_5$-coincident NS5- or M5-branes, i.e., fixed $N_5$ and $N_2\to\infty$. In this case, it is expected that the eigenvalue distribution of $Y_4,\cdots,Y_9$ becomes five-sphere that corresponds to spherical NS5- or M5-branes~\cite{Asano:2017nxw,Asano:2017xiy}, modulo the assumption that $Y_4,\cdots,Y_9$ are simultaneously diagonalizable. 
If this expectation is true, then the geometric partitioning on five-branes can be related to the partitioning of matrix degrees of freedom. 
Unlike the partitioning of fuzzy sphere (Fig.~\ref{fig:Fuzzy_sphere}), 
in this case it is not immediately clear how to treat the off-diagonal entries. Both \eqref{matrix-partitioning-A-Abar} and \eqref{matrix-partitioning-A-B-C} seem to be natural. 

\section{Application 2: Small black hole and Hawking radiations}\label{sec:BH-evaporation}
In this section, we consider small black hole and Hawking radiations as an application of the matrix entanglement. 

As shown by Horowitz~\cite{Horowitz:1999uv}, a small black hole in AdS$_5\times$S$^5$ described by dual 4d SYM on S$^3$ can evaporate into gas of gravitons if the coupling constant on the QFT side is sufficiently large ($\lambda_{\rm 't Hooft}\gg N^{8/17}$)~\footnote{
For the stringy corrections to type IIB supergravity to be small, $g_{\rm YM}^2\sim g_s\ll 1$ and $\frac{\alpha'}{R_{\rm AdS}^2}\sim(g_{\rm YM}^2N)^{-1/2}\ll 1$ are required. 
In the 't Hooft large-$N$ limit ('t Hooft coupling $\lambda=g_{\rm YM}^2N$ fixed), the former condition is automatically satisfied, and the latter can be satisfied as well if $\lambda$ is taken to be large. 
To realize $\lambda_{\rm 't Hooft}\gg N^{8/17}$ without having large stringy corrections, we should take $g_{\rm YM}^2\sim N^{-\alpha}$, $0\le\alpha\le\frac{9}{17}$.
}$^,$\footnote{
If the energy of black hole is not sufficiently small, graviton gas becomes dense before black hole completely evaporates and the equilibrium state consisting of black hole and graviton gas is realized. If $\lambda_{\rm 't Hooft}\lesssim N^{8/17}$, the stringy effect is not negligible for such a small black hole. This is the reason that $\lambda_{\rm 't Hooft}\gg N^{8/17}$ is required.  
} and energy is sufficiently small ($E\ll N^{20/17}$).
The same might be true for 11d black hole described by Matrix Theory~\cite{Banks:1997tn,Banks:1997cm,Horowitz:1997fr,Bergner:2021goh}. 
Note that there is no need for external heat bath, and hence such systems provide us with a theoretically clean setup to study the black hole information problem.  

As discussed in Sec.~\ref{sec:partially-deconfined-state}, small black hole is described by partially-deconfined states. 
A simple mechanism of the emergence of Page curve naturally follows from partial deconfinement.
(See Ref.~\cite{Krishnan:2020oun} for an early suggestion that partial deconfinement should be important to resolve the information puzzle.)
Below, we first consider a small but not too small black hole that does not evaporate (Sec.~\ref{sec:Bekenstein-Hawking}), and then move on to the case of sufficiently small black hole that evaporates completely (Sec.~\ref{sec:Island}).
\subsection{Bekenstein-Hawking entropy from entanglement}\label{sec:Bekenstein-Hawking}
Let us consider a small black hole without radiations\footnote{
More precisely, we consider a situation that black hole does not evaporate completely and the density of radiations is negligibly small. 
} that is represented by a superposition of wave packets of the form \eqref{partially-deconfined-state}. 
We can define the reduced density matrix and von Neumann entropy $S_A$ by keeping only the deconfined sector (region $A$), as in \eqref{MM-entanglement-def-1}. 
We would like to know whether $S_A$ obtained in this way can be the Bekenstein-Hawking entropy. 

Let $N_1$ be the size of the deconfined sector, $N_2=N-N_1\gg N_1$, and the coarse-grained entropy be  
\begin{align}
S_{\rm coarse\ grained}
=
\log K. 
\end{align} 
Then, there are $K$ wave packets $|Y_{[a]},Q_{[a]}\rangle$ ($a=1,2,\cdots,K$) that give the same values for the macroscopic observables (energy, coarse-grained eigenvalue distribution, etc) up to $1/N$-corrections and the coarse-grained reduced density matrix is written as
\begin{align}
\hat{\rho}_{\rm coarse\ grained}
=
\frac{1}{K}|Y_{[a]},Q_{[a]}\rangle\langle Y_{[a]},Q_{[a]}|. 
\end{align} 
Only the upper-left block (region $A$) of $(Y_{[a]},Q_{[a]})$, denoted by $(Y^{(1)}_{[a]},Q^{(1)}_{[a]})$, is nonzero, as in \eqref{partially-deconfined-state}. 
At weak coupling, explicit computations show $K\sim e^{N_1^2}$~\cite{Hanada:2019czd}. 
On the gravity side, this coarse-grained entropy should be the Bekenstein-Hawking entropy. 
Therefore, our task is to see whether this coarse-grained entropy coincides with the von Neumann entropy $S_A$. 

We consider a superposition of $K$ wave packets as a typical state corresponding to the small black hole: 
\begin{align}
|\Phi\rangle
\sim
\frac{1}{\sqrt{K}}\sum_{a=1}^K
|Y_{[a]},Q_{[a]}\rangle. 
\end{align}
We would like to claim that, 
by tracing out $\mathcal{H}_{\bar{A}}$, we obtain
\begin{align}
{\rm Tr}_{\bar{A}}
\left(|\Phi\rangle\langle\Phi|\right)
\sim
\frac{1}{K}\sum_{a=1}^K
|Y_{[a]}^{(1)},Q_{[a]}^{(1)}\rangle
\langle Y_{[a]}^{(1)},Q_{[a]}^{(1)}|.
\label{eq:BH_entropy_from_entanglement} 
\end{align}
Then the von Neumann entropy is also $S_A\sim\log K$. 
Intuitively, we would like to claim that region $\bar{A}$ behaves like a heat bath. 

For \eqref{eq:BH_entropy_from_entanglement} to hold, 
\begin{align}
\frac{1}{K}
\sum_{a\neq b}
{\rm Tr}_{\bar{A}}
\left(
|Y_{[a]},Q_{[a]}\rangle
\langle Y_{[b]},Q_{[b]}|
\right)
\end{align}
has to be negligible. One might think this is trivial because the overlap between low-energy wave packets are exponentially suppressed at large $N$. 
However we have to be a little bit careful here: we have to confirm that the overlap is small in ${\cal H}_{\bar{A}}$, where $Y$ and $Q$ are zero. 
We argue that this is indeed the case, and 
\begin{align}
{\rm Tr}_{\bar{A}}
\left(
|Y_{[a]},Q_{[a]}\rangle
\langle Y_{[b]},Q_{[b]}|
\right)
\sim
c\cdot
|Y_{[a]}^{(1)},Q_{[a]}^{(1)}\rangle
\langle Y_{[b]}^{(1)},Q_{[b]}^{(1)}|
\end{align}
with
\begin{align}
c
\lesssim
e^{-N_1N_2}. 
\end{align}
To motivate this scaling, let us consider a simplified situation where $Y^{(1)}_{[a]}$ and $Y^{(1)}_{[b]}$ are diagonal, $Q^{(1)}_{[a]}=Q^{(1)}_{[b]}=0$, and eigenvalues are sufficiently far separated such that off-diagonal entries can be treated as harmonic oscillators with frequency $\omega$ proportional to the distance between eigenvalues.
Let $\vec{y}_{[a]}^i$ and $\vec{y}_{[b]}^i$ be the nine-vectors representing $(i,i)$-components of $Y_{[a]}$ and $Y_{[b]}$. 
For the $(i,j)$-component, $\omega=g|\vec{y}_{[a]}^i-\vec{y}_{[a]}^j|$ or $g|\vec{y}_{[b]}^i-\vec{y}_{[b]}^j|$; let's take this to be large but of order $N^0$.
For $i=1,\cdots,N_1$ and $j=N_1+1,\cdots,N$, the wave functions are the normal distribution with the width $(g\cdot|\vec{y}_{[a]}^i|)^{-1/2}$ and $(g\cdot|\vec{y}_{[b]}^i|)^{-1/2}$, respectively. 
Then, unless $|\vec{y}_{[a]}^i|=|\vec{y}_{[b]}^i|$ (which is not the case in general), the wave functions do not overlap perfectly, and the inner product of these components in two wave packets are suppressed.
There are $N_1N_2$ matrix degrees of freedom contributing to such a supression, and hence the exponentially small suppression factor $e^{-N_1N_2}$ is expected. 
The lower-right block ($i,j=N_1+1,\cdots,N$) may give further suppressions when they are traced out. We expect similar suppression also when $Y_{[a]}$ and $Y_{[b]}$ are not simultaneously diagonalizable, because there is no reason to expect perfect overlap of wave function along any of the dimensions.  
Therefore, \eqref{eq:BH_entropy_from_entanglement} should hold, and $S_A=\log K$. 

\begin{figure}[htbp]
  \begin{center}
   \includegraphics[width=60mm]{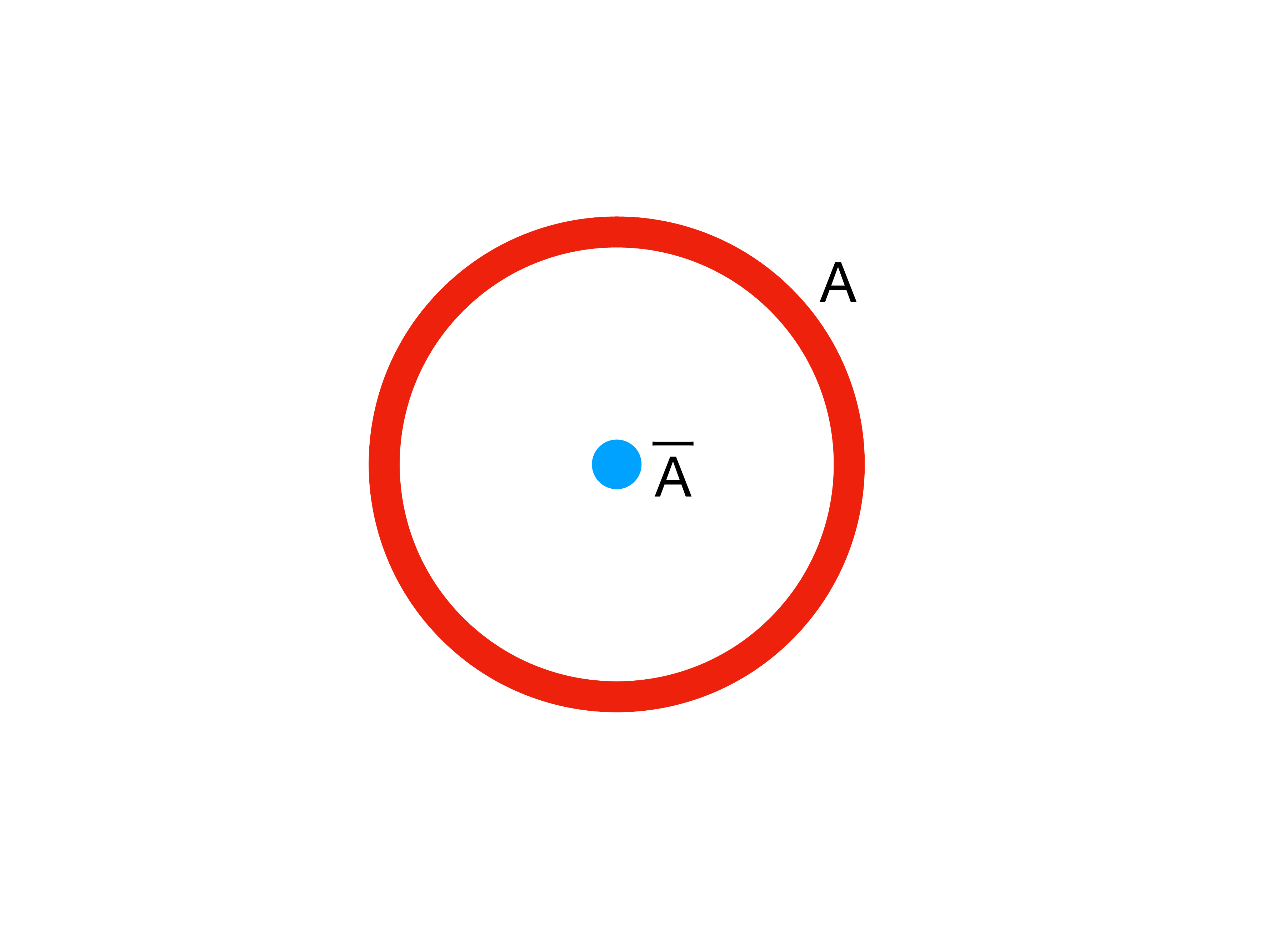}
  \end{center}
  \caption{
  Geometry of a small black hole encoded into a partially-deconfined state, in terms of D-brane configuration. 
    Region $A$ describes the degrees of freedom near the black hole horizon. 
  D-branes described by region $\bar{A}$ is sitting at the origin of $\mathbb{R}^{9-p}$. 
  Only region $A$ contributes to the coarse-grained entropy. 
  }\label{fig:Island-geometry-1}
\end{figure}

So far we called region $A$ in this picture as `black hole', but in fact, D-branes described by region $\bar{A}$ are also sitting inside the `black hole'; they are sitting at the origin of the bulk geometry. (See Sec.~\ref{sec:partially-deconfined-state}.)
It seems to be natural to interpret region $A$ as the degrees of freedom near the black hole horizon as depicted in Fig.~\ref{fig:Island-geometry-1}, 
because the deconfined sector describes an extended bound state of D-branes and strings, and hence, D-branes in region $A$ cannot be sitting at the origin. 
  
In terms of matrix degrees of freedom, the formation process of black hole would look like Fig.~\ref{fig:BH-formation-evaporation}, from left to right. In the figure, first there were multiple small objects ($\sim$ multiple small deconfined sectors). Eventually, they merge and form a single deconfined sector. 
When a black hole is newly formed, we do not expect that regions $A$ and $\bar{A}$ are fully entangled.
As an extreme example, we can imagine a pure state consisting of just one specific wave packet $|Y,Q\rangle$. 
After some time (presumably the scrambling time), the system evolves to a generic superposition of many wave packets and then $A$ and $\bar{A}$ are strongly entangled. The argument above assumes such a situation. 

\subsubsection*{Two stacks of D-branes}
As an analogous but slightly different situation, let us consider two stacks of D-branes considered in Sec.~\ref{sec:2-BH-TFD}. 
We assume that $Y_I$'s and $Q_I$'s are block-diagonal as in \eqref{block-diag}.  
To be concrete, we assume each block is $N_1=N_2=\frac{N}{2}$. 
Furthermore, we assume that $Y_1$ is close to ${\rm diag}\left(\frac{L}{2},\cdots,\frac{L}{2},-\frac{L}{2},\cdots,-\frac{L}{2}\right)$, where $+\frac{L}{2}$ and $-\frac{L}{2}$ appear $\frac{N}{2}$ times for each, and other $Y_I$'s and $Q_I$'s are close to zero. 
Still, we can consider a linear combination of wave packets of this form:
\begin{align}
|\Phi\rangle
=
\sum_a c_a|Y_{[a]},Q_{[a]}\rangle. 
\end{align}
In the limit of $L\to\infty$, off-diagonal blocks decouple. Thermofield double states are special case of such two-stack states. 
It is convenient to trace out such decoupled degrees of freedom (region $C$ in \eqref{matrix-partitioning-A-B-C}) and keep diagonal blocks (region $A$ and region $B$ in \eqref{matrix-partitioning-A-B-C}). In this case, matrix entries in the off-diagonal blocks are harmonic oscillators with frequency $\omega=gL$ up to negligible corrections, and hence 
they are indistinguishable regardless of the details of the diagonal block, unlike the case of the small black hole considered above. 
(Strictly speaking, we are considering $L\to\infty$ for large but fixed $N$.)
We obtain 
\begin{align}
\hat{\rho}_{A\cup B}
&=
{\rm Tr}_{\mathcal{H}_C}
|\Phi\rangle\langle\Phi|
\nonumber\\
&=
\sum_{a,b} c_ac_b^\ast
|Y_{[a]}^{(1)},Q_{[a]}^{(1)}\rangle
|Y_{[a]}^{(2)},Q_{[a]}^{(2)}\rangle
\langle Y_{[b]}^{(1)},Q_{[b]}^{(1)}|
\langle Y_{[b]}^{(2)},Q_{[b]}^{(2)}|
\nonumber\\
&=
\left(
\sum_{a} c_a
|Y_{[a]}^{(1)},Q_{[a]}^{(1)}\rangle
|Y_{[a]}^{(2)},Q_{[a]}^{(2)}\rangle
\right)
\left(
\sum_{b} 
c_b^\ast
\langle Y_{[b]}^{(1)},Q_{[b]}^{(1)}|
\langle Y_{[b]}^{(2)},Q_{[b]}^{(2)}|
\right). 
\end{align}
The decoupled degrees of freedom do not play nontrivial role in entanglement. 

\subsection{Evaporation and Page curve}\label{sec:Island}
\begin{figure}[htbp]
  \begin{center}
   \includegraphics[width=120mm]{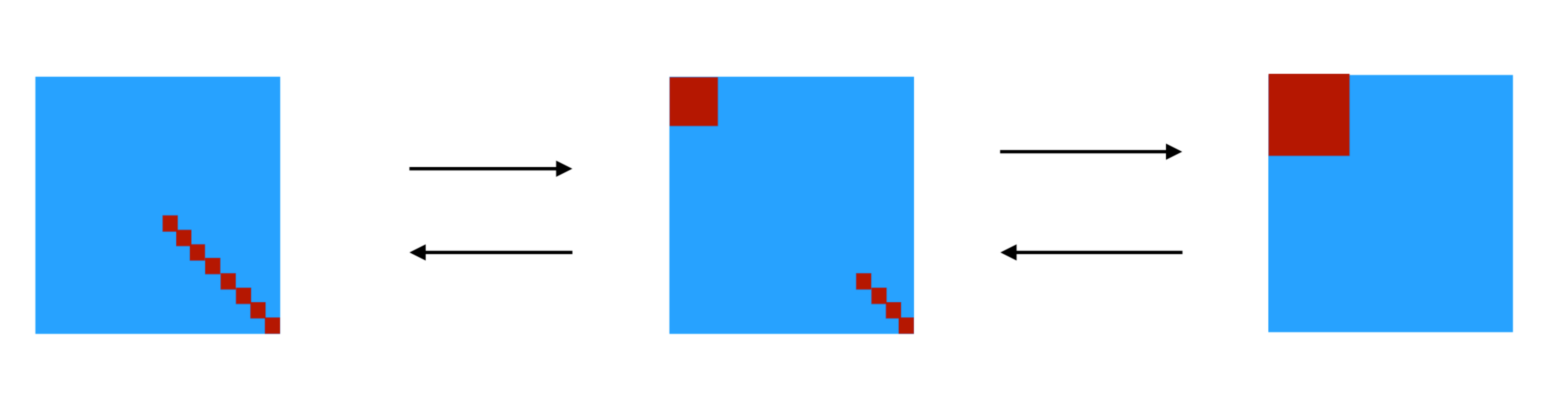}
  \end{center}
  \caption{
  A cartoon picture of the formation (left to right) and evaporation (right to left) of small black hole encoded into matrix degrees of freedom. 
  }\label{fig:BH-formation-evaporation}
\end{figure}

A cartoon picture of the evaporation of small black hole in terms of matrix degrees of freedom is shown in Fig.~\ref{fig:BH-formation-evaporation}, from right to left. Different from the black hole formation (from left to right), small blocks should correspond to gravitons; see Sec.~\ref{sec:partially-deconfined-state}.

\begin{figure}[htbp]
  \begin{center}
    \includegraphics[width=120mm]{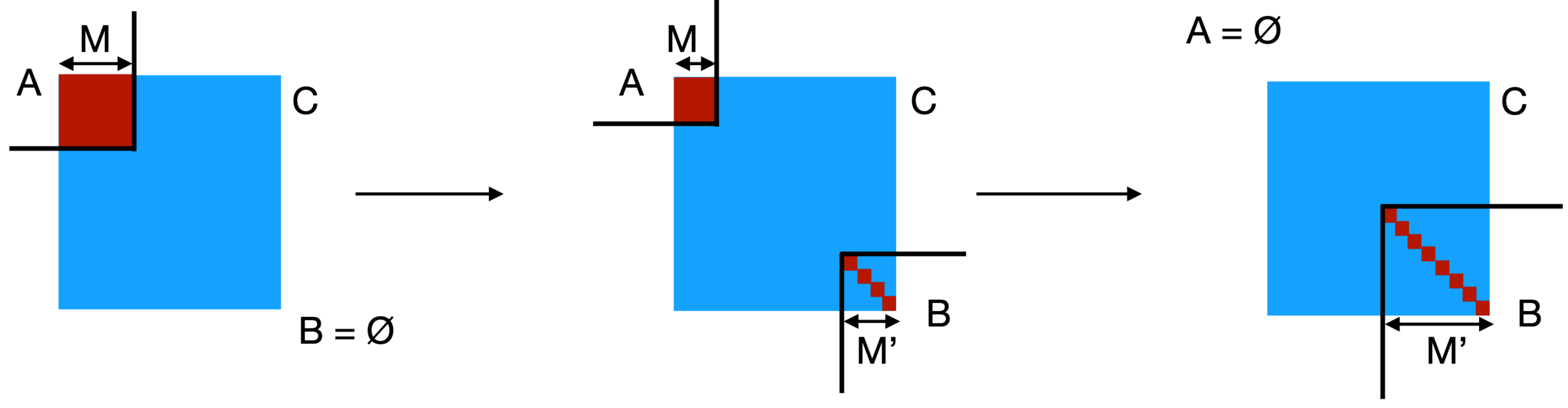}
  \end{center}
  \caption{
  Small black hole (region $A$), radiations (region $B$) and the confined sector (region $C$) in terms of matrix degrees of freedom. 
  (There is an ambiguity regarding whether the off-diagonal elements (strings between D-branes) should be included in region $B$.)
  }\label{fig:island?}
\end{figure}

  Suppose that the small BH corresponds to $M\times M$ deconfined block (region $A$) and radiations are described by 
  $M'\times M'$ block (region $B$). In addition, there is a confined sector (region $C$); see Fig.~\ref{fig:island?}. 
  We assume that $M,M'\ll N$, because otherwise the black hole cannot evaporate completely. 
  We trace out the Hilbert space corresponding to the confined sector $C$, and define the reduced density matrix $\hat{\rho}_{A\cup B}$:
\begin{align}
\hat{\rho}
=
\hat{\rho}_{A\cup B\cup C}, 
\qquad
\hat{\rho}_{A\cup B}
=
{\rm Tr}_C\hat{\rho}. 
\end{align}
  
\begin{figure}[htbp]
  \begin{center}
     \includegraphics[width=60mm]{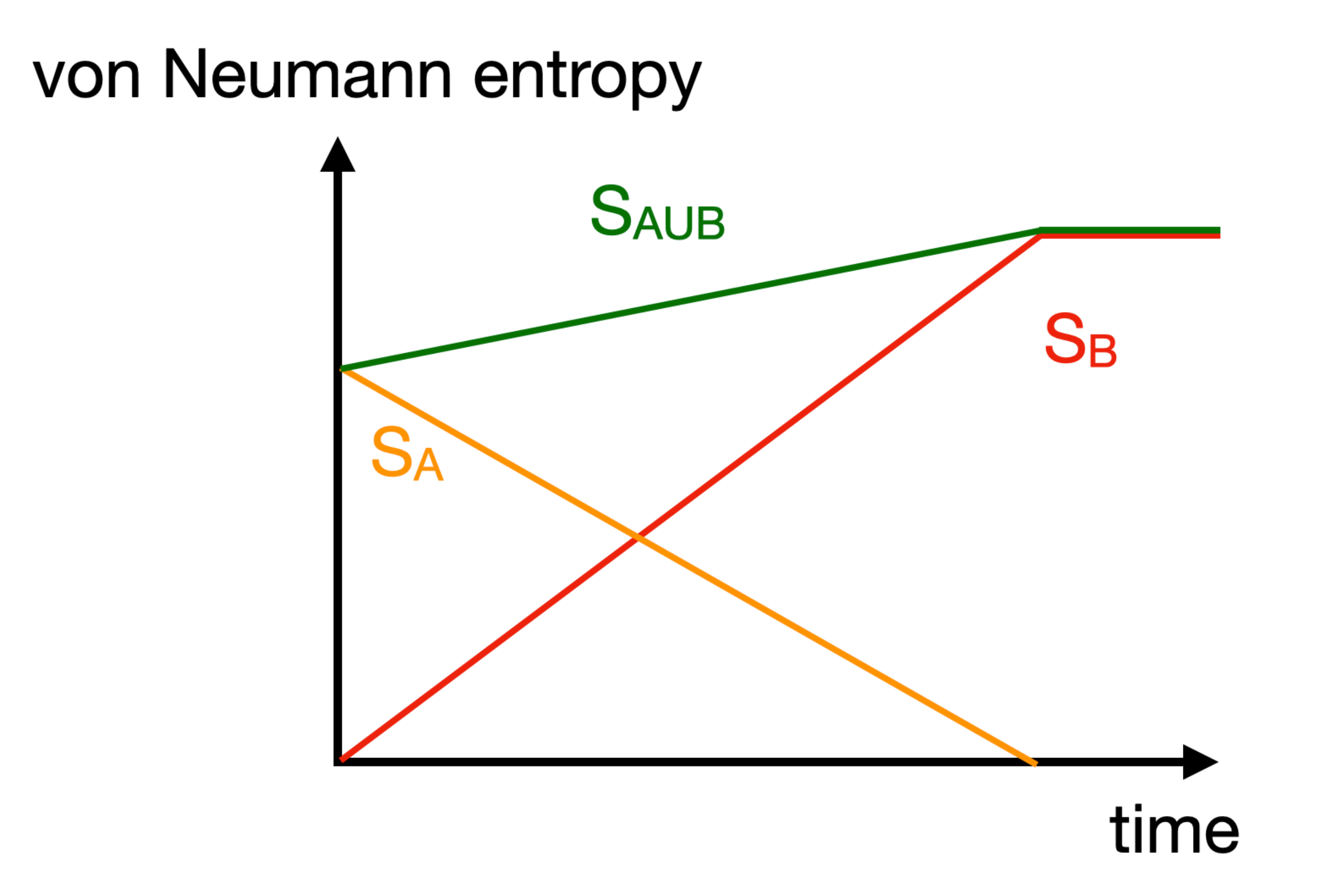}
  \end{center}
  \caption{
  Qualitative behavior of $S_A$ (entropy of black hole), $S_B$ (entropy of radiations) and $S_{A\cup B}$. Apparent departure from the Page curve can be cured by taking into account the confined sector (region $C$); see Fig.~\ref{fig:island?}.
  }\label{fig:EE-BH-radiation}
\end{figure}

Reduced density matrix $\hat{\rho}_{A\cup B}$ obtained this way is a mixed state, even if the original density matrix $\hat{\rho}$ is pure.
Hence the von Neumann entropy $S_{A\cup B}$ (entropy of the system of black hole and radiations)
  \begin{align}
  S_{A\cup B}=-{\rm Tr}_{A\cup B}\hat{\rho}_{A\cup B}\log\hat{\rho}_{A\cup B}
  \end{align} 
is nonzero. 
We can also define the von Neumann entropy $S_A$ (entropy of black hole) and $S_B$ (entropy of radiations) as
  \begin{align}
  & S_A=-{\rm Tr}_A\hat{\rho}_A\log\hat{\rho}_A,
  \qquad 
  \hat{\rho}_A={\rm Tr}_B\hat{\rho}_{A\cup B}, 
  \nonumber\\
  & S_B=-{\rm Tr}_B\hat{\rho}_B\log\hat{\rho}_B, 
  \qquad
  \hat{\rho}_B={\rm Tr}_A\hat{\rho}_{A\cup B}. 
  \end{align} 
Based on the argument along the lines of Sec.~\ref{sec:Bekenstein-Hawking}, these von Neumann entropies should be the same as the coarse-grained entropy.    
As the black hole evaporates, $S_A$ decreases, while $S_B$ increases.
The crucial point is that $M$ and $M'$ change with time; see Fig.~\ref{fig:EE-BH-radiation}.  
Note that $S_{A\cup B}=S_A$ at early time and $S_{A\cup B}=S_B$ at late time. 
There is no reason to expect that region $A$ and region $B$ are directly entangled, and hence we expect $S_{A\cup B}=S_A+S_B$.  

Apparently, the time dependence of $S_{A\cup B}$ does not exhibit the Page curve. As we will see, however, there is no contradiction. One has to take into account the degrees of freedom in the confined sector to obtain the Page curve. 
This is different from the assumption in Page's original argument~\cite{Page:1993wv} that black hole and radiations form a pure state. 

\subsubsection{Page curve}\label{sec:Page_curve}
Let us introduce regions $A,B$ and $C$ as shown in Fig.~\ref{fig:island?}. 
The region $C$ corresponds to $Y=Q=0$ and hence it does not contribute to the coarse-grained entropy. 
If we trace out region $C$, we obtain the reduced density matrix discussed above.
Note that $C$ contains most of the matrix degrees of freedom; otherwise the small black hole cannot evaporate completely. 
In Fig.~\ref{fig:island?}, $A$ and $B$ are drawn much bigger than their actual sizes. 

Because the entire system $A\cup B\cup C$ is in a pure state, we immediately obtain
\begin{align}
S_{A\cup C} = S_B
\end{align} 
and
\begin{align}
S_{B\cup C} = S_A. 
\label{eq:entanglement-island}
\end{align} 
Therefore, if we interpret that ${\rm min}(S_B,S_{B\cup C})$ is the entropy of radiations, we obtain the Page curve; see Fig.~\ref{fig:Page-curve}. 

Clearly, there is a close similarity to the island formula~\cite{Penington:2019npb,Almheiri:2019psf}:
it appears that two ways of partitioning the matrix degrees of freedom ($A$ and $B\cup C$, or $A\cup C$ and $B$) correspond to two ways of choosing the extremal surfaces in the island formula. 
In Sec.~\ref{sec:island-or-not}, we will discuss whether this mechanism is related to the island proposal or not. At this stage, we will study the matrix entanglement as it is, without referring to the island proposal. 

\subsubsection{Meaning of the confined sector on the gravity side}\label{sec:gravity_side_interpretation}

\begin{figure}[htbp]
  \begin{center}
   \includegraphics[width=60mm]{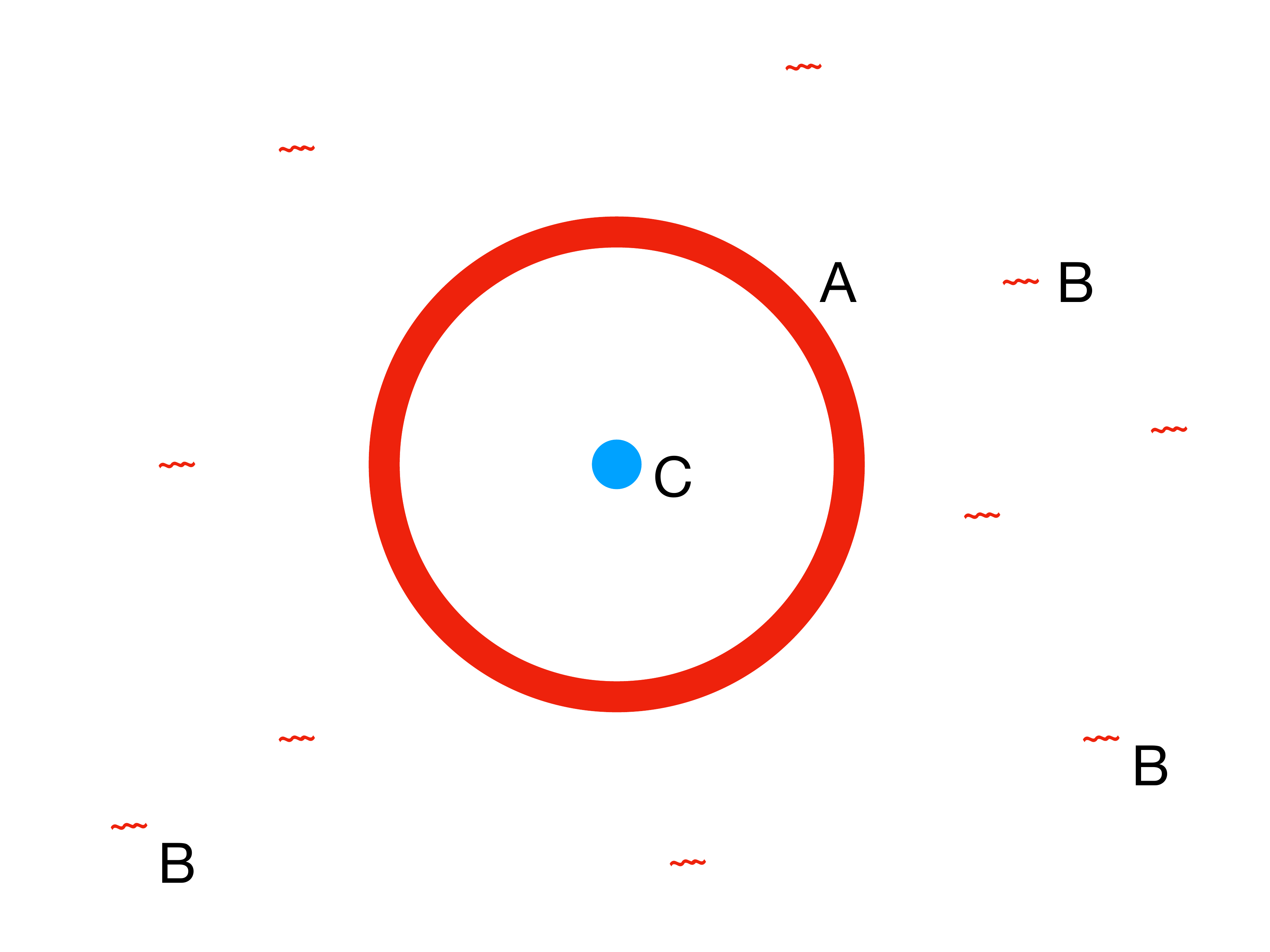}
  \end{center}
  \caption{
  Region $A$ describes the degrees of freedom near the black hole horizon. Region $B$ describes Hawking radiations. 
  D-branes described by region $C$ are sitting at the origin of $\mathbb{R}^{9-p}$, inside the horizon. 
  }\label{fig:Island-geometry-2}
\end{figure}

Geometry obtained by identifying matrix eigenvalues and D-brane locations~\cite{Hanada:2021ipb,Hanada:2021swb,Witten:1995im} is shown in Fig.~\ref{fig:Island-geometry-2}.
(Here we assumed that the radial coordinates determined from small blocks are large such that they can be regarded as gravitons outside the black hole. Explicit demonstration of this picture is desirable to strengthen our claim below.)  
Region $A$ describes the degrees of freedom near the black hole horizon. 
D-branes in region $C$ are sitting behind the horizon. Region $C$ is strongly entangled with radiations (region $B$) at late time, and when this region and radiations are combined, the von-Neumann entropy becomes much smaller, as shown in Fig.~\ref{fig:Page-curve}. 

In Page's argument, only radiations and black hole are concerned. Therefore, if we want to put our finding to his framework, we have to include the region $C$ as a part of radiations or black hole. 
Time dependence of von Neumann entropy shown in Fig.~\ref{fig:Page-curve} suggests that, before and after the Page time (the time when $S_A=S_B$ holds), $B$ and $B\cup C$ should be regarded as ``radiations", respectively.
Can we justify this interpretation?

It is easy to see that the alternative choice leads to a pathology. 
We assume that $S_A$ and $S_B$ agree with the coarse-grained entropy of black hole and radiations, respectively. 
(This is the reason that we believe that black hole can completely evaporate, based on the calculation on the gravity side~\cite{Horowitz:1999uv}.)
Intuitively, and somewhat naively, this is the number of independent wave packets, i.e., possible choice of $Y$ and $Q$, in each sector.
Seen this way, region $C$ does not contribute to the coarse-grained entropy, because $Y$ and $Q$ vanish there. 
However if we use precise definition of coarse-grained entropy (i.e., consider all possible density matrices that give the same expectation values for macroscopic observables under consideration, calculate their von Neumann entropies, and then choose the largest value as `coarse-grained entropy'; see e.g., Ref.~\cite{Almheiri:2020cfm}), region $C$ does have a visible contribution. 
Suppose we interpreted that $B\cup C$ describes radiations before Page time. Then the von Neumann entropy $S_{B\cup C}=S_A$ is larger than $S_B$. Then the coarse-grained entropy cannot be $S_B$, because the von Neumann entropy gives the lower bound for the coarse-grained entropy. 
We would encounter the same pathology if we interpreted that $A\cup C$ describes black hole (equivalently, $B$ describes radiations) after the Page time, because the von Neumann entropy $S_{A\cup C}=S_B$ is larger than $S_A$ which we want to have as the coarse-grained entropy. 
Therefore, as long as we split the system to two pieces (``black hole" and ``radiations") in a way consistent with semiclassical picture on the gravity side,
we are forced to interpret that ``Hilbert space of radiations" is $\mathcal{H}_B$ before the Page time and $\mathcal{H}_{B\cup C}$ after the Page time.
It would mean that, although both orange and red lines in Fig.~\ref{fig:Page-curve} are well-defined on the gauge theory side, the dual gravity description in semiclassical language can be valid only for the gray line (Page curve). 

To have a better argument, we have to specify the rule of the game precisely: which part of the Hilbert space can Bob, the exterior observer, manipulate? Clearly, we should assume that Bob can access to the region $B$, but not to region $A$.  
How about the confined sector (region $C$)? Although D-branes in the confined sector is sitting at the origin and hence they are behind the horizon of the small black hole at the center of the bulk, they interact with region $B$ via off-diagonal entries (open strings), and due to such interactions, radiations in region $B$ know they are in asymptotically AdS$_5\times$S$^5$ spacetime. Furthermore, when the AdS radius is very large and black hole is very small, black hole can sit away from the origin at least for some time, if we fine-tune the initial condition.\footnote{
There are two effects which pull black hole toward the center. Firstly, there is a conformal mass term for scalar fields.
Secondly, D-branes attract with each other when supersymmetry is broken by thermal excitations.
The latter works in the BFSS matrix model as well. 
}
It is as if region $C$ prepares the background spacetime which black hole and radiations  live in. See Sec.~\ref{sec:discussion} for further discussion. Therefore, we find it more natural to assume that Bob has access to region $C$ as well. 
In the following, we assume that Bob can access $B\cup C$. Specifically, we assume that Bob knows the reduced density matrix $\hat{\rho}_{B\cup C}$. 

As usual, Bob's task is to read Alice's diary which was thrown into the black hole. For that purpose, he wants to know as much information on radiations as possible. He has three options: (1) keep $\hat{\rho}_{B\cup C}$, (2) discard $C$ and get $\hat{\rho}_B$, and (3) discard $B$ and get $\hat{\rho}_C$. He wants to specify the state as precisely as possible, and hence, he wants to minimize the entropy. The von Neumann entropies associated with (1), (2) and (3) are $S_{B\cup C}=S_A$, $S_B$ and $S_C=S_A+S_B$, respectively. Therefore, before and after the Page time, he should choose option (2) and option (1), respectively. The Page curve represents the best of Bob's knowledge.

\subsubsection{Time scales associated with the evaporation}\label{sec:time_scale}
Before the Page time, the smallest von Neumann entropy Bob can achieve is $S_B$, that is the coarse-grained entropy of radiations. He does not know the information on the confined sector.~\footnote{He can decrease the entropy further by throwing away a part of region $B$, but that does not add any information on $C$.
}$^,$\footnote{
At weak coupling, entanglement between region $A$ and region $C$ (resp., $B$ and region $C$) is encoded into the off-diagonal blocks corresponding to open strings between D-branes in region $A$ and region $C$ (resp., $B$ and region $C$). By specifically looking at those off-diagonal blocks, we can extract the information regarding black hole or radiations. However, at strong coupling, it is highly unlikely that such a simple partitioning exists. Note that the gauge invariance of the confined sector restricts the possible options for partitioning rather strictly. 
}
On the other hand, after the Page time he can make the entropy as small as $S_{B\cup C}=S_A<S_B$. The confined sector gave him some information. 
In this sense, Bob can ``see" the region $C$ only after the Page time. 

\begin{figure}[htbp]
  \begin{center}
    \includegraphics[width=60mm]{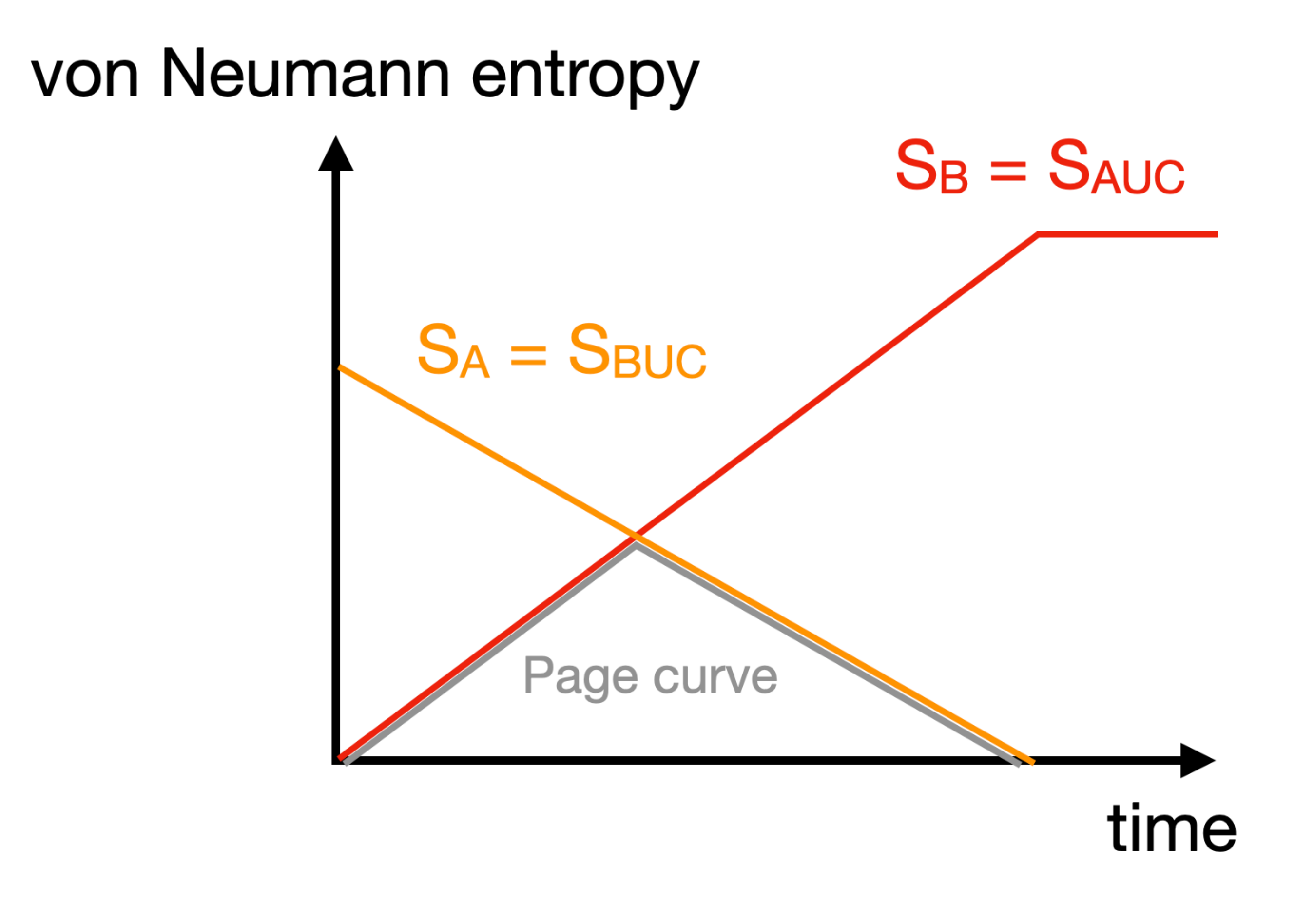}
  \end{center}
  \caption{ Regions $A$ and $B$ are interpreted as black hole and radiations, respectively.
  Region $C$ describes D-branes sitting at the origin, which is inside the horizon. After the Page time, it is natural to interpret that $C$ is a part of radiations. 
    }\label{fig:Page-curve}
\end{figure}

It is important to note that the estimate of von Neumann entropy above is valid only after certain time needed for regions $A$, $B$ and $C$ to fully entangle with each other. 
When a black hole is newly formed, we do not expect that region $C$ is fully entangled with region $A$. Region $B$ does not exist or is very small at this point. As an extreme example, we can imagine a pure state consisting of just one specific wave packet $|Y,Q\rangle$. 
After some time, the system evolves into a generic superposition and then regions $A$, $B$ and region $C$ are strongly entangled. Due to such entanglement, $S_A$ and $S_B$ should coincide with the coarse-grained entropies of black hole and Hawking radiation, respectively. The time dependence of the entropies shown in Fig.~\ref{fig:Page-curve} applies after this time scale.
Such a time dependence closely resemble the time scale for the formation of the island on the gravity side. 

To see when Bob can steal Alice's secret, let us slightly modify the Hayden-Preskill information retrieval protocol~\cite{Hayden:2007cs}.  
The original Hayden-Preskill protocol assumes maximal entanglement between black hole and a subsystem of radiations after the Page time. This assumption is satisfied if $\mathcal{H}_{B\cup C}$ is regarded as ``Hilbert space of radiations" after the Page time. Suppose Alice throw her diary (a diagonal block in region $B$) into black hole (region $A$) after the Page time.
Then region $A$ becomes slightly bigger, but region $C$ is not affected instantaneously. A natural time scale when the information thrown into region $A$ is transferred to region $C$ is the scrambling time.
Although this information transfer can take place even before the Page time, Bob can have the information on the region $C$ only after the Page time. 
Therefore, he can read Alice's diary only after the Page time.

\subsubsection{Entanglement Island?}\label{sec:island-or-not}
In the discussions regarding the Page curve and confined sector, there were several features of the confined sector (region $C$) that closely resemble the entanglement island:
\begin{itemize}
\item
D-branes in the confined sector are sitting behind the horizon (Sec.~\ref{sec:gravity_side_interpretation} and Fig.~\ref{fig:Island-geometry-2}).

\item
Combined with radiations (region $B$), it reproduces the entropy of black hole (region $A$), i.e.,~$S_{A}=S_{B\cup C}$ (Sec.~\ref{sec:Page_curve}).

\item
Bob can see the confined sector only after the Page time (Sec.~\ref{sec:time_scale}). 

\item
Time scale for the information retrieval appears to be the same (Sec.~\ref{sec:time_scale}). 

\end{itemize}
Therefore, if the gravity side admits the entanglement island, it would be natural to expect that the confined sector corresponds to the island. 

Still, there are several points that appear to be different from the entanglement island proposal. Firstly, it may not be appropriate to interpret that the confined sector describes the interior of black hole. Rather, it may describe the AdS$_5\times$S$^5$ spacetime that the black hole and radiations live in; see a discussion in Sec.~\ref{sec:discussion}. 
Secondly, the notion of the entanglement wedge, which is important in the island proposal, is not clear in terms of matrix degrees of freedom. Note also that there is a debate concerning the island proposal when the non-gravitational heat bath is not introduced.    
Because our setup does not involve non-gravitational heat bath, we need to introduce the interface separating black hole and radiations at such a location where gravity is dynamical. There are arguments against~\cite{Raju:2020smc,Geng:2020fxl,Raju:2021lwh} and supporting~\cite{Ghosh:2021axl,Krishnan:2021ffb} the use of the holographic entanglement formula and the existence of the entanglement island in such a case. 
Although our argument on the gauge theory side is not affected by this debate, the dual gravitational interpretation would be different depending on the conclusion of the debate. There is also an argument against existence of islands in a theory of massless gravity~\cite{Geng:2021hlu}.  Perhaps matrix entanglement can provide us with a concrete setup to better understand this issue.\footnote{We thank Hao Geng, Chethan Krishnan, Suvrat Raju and Christoph Uhlemann for useful comments regarding this issue.} 

It would be fair to say that we do not know the precise relationship between the confined sector in gauge theory and the entanglement island in gravity. Whether they are the same or not, we expect that matrix entanglement is one of the keys to the understanding of the black hole information problem in gauge/gravity duality. 
\subsection{Type IIA region of D0-brane matrix model (Toy model for black hole evaporation)}\label{sec:zero-brane-evaporation}
The small black hole discussed above was conceptually very clean on the gravity side. However, analytic calculations on the QFT side was difficult. 
Below, let us show a toy model~\cite{Berkowitz:2016znt,Berkowitz:2016muc} that has qualitatively similar behavior from QFT point of view, 
although the weakly-coupled dual gravity description is lacking. 

\begin{figure}[htbp]
  \begin{center}
   \includegraphics[width=100mm]{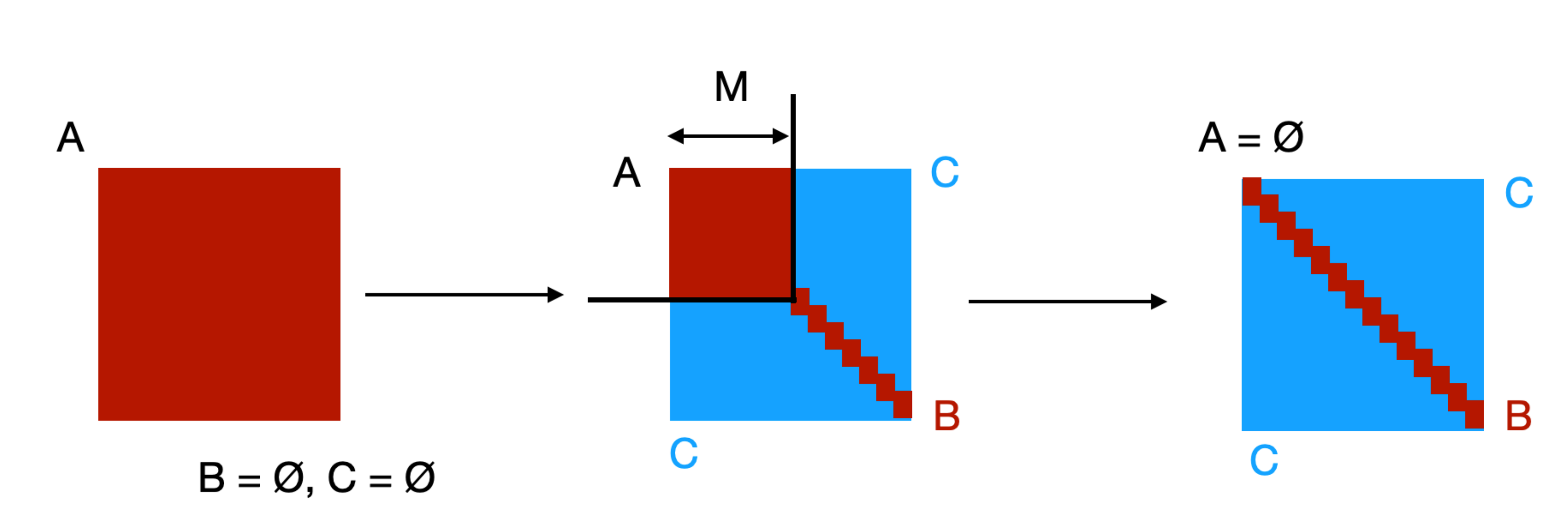}
   \includegraphics[width=60mm]{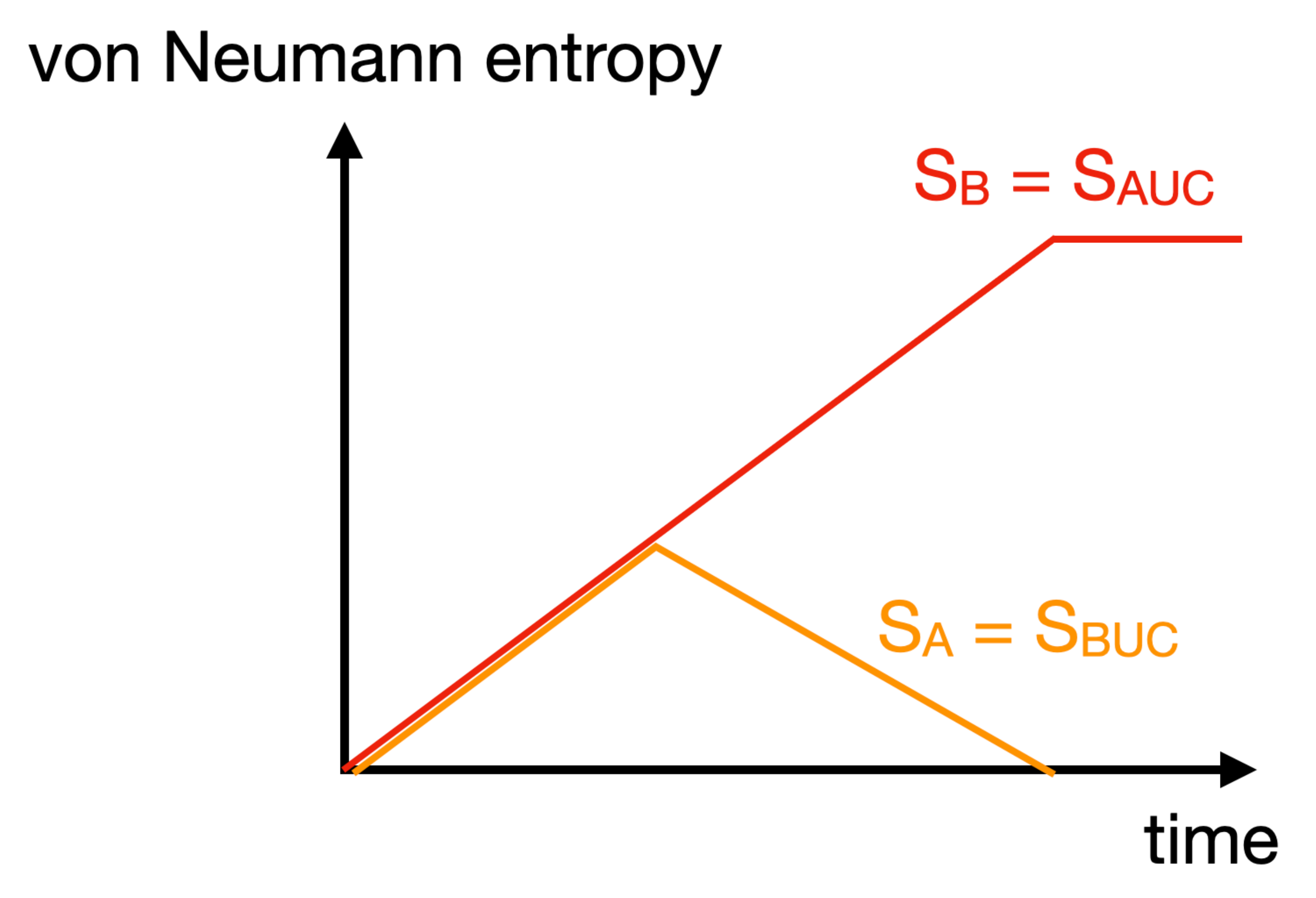}
  \end{center}
  \caption{
Type IIA black zero-brane described by D0-brane matrix model (region $A$) can `evaporate' by emitting D0-branes (region $B$) one by one. 
Black zero-brane is described by the $M\times M$-block, where $M$ decreases from $N$ to $0$. 
At late time, off-diagonal elements shown by blue (region $C$) decouple from the dynamics. However, they affect the calculation of the von Neumann entropy.
  }\label{fig:IIA-D0-evaporation}
\end{figure}

Let us consider the parameter region of D0-brane matrix model that is dual to type IIA black zero-brane. The black zero-brane is a bound state of many D0-branes and open strings. 
As the initial condition, let us assume that all $N$ D0-branes described by the U($N$) theory are in the black zero-brane. 
At large but finite $N$, the black zero-brane can decay by emitting D0-branes toward flat direction at a very long time scale. (The time scale for the emission of first few D0-branes is of order $e^N$.) 
A cartoon picture of such `evaporation' is shown in Fig.~\ref{fig:IIA-D0-evaporation}. 
Emitted D0-branes will run away to infinity. They are analogous to the Hawking radiations as far as the eigenvalues are concerned, although they are not massless. 
The black zero-brane is described by the $M\times M$-block, where $M$ decreases from $N$ to $0$. 
As D0-branes (diagonal entries) escape to infinity, open strings (off-diagonal entries) become infinitely heavy and decouple from the dynamics. In terms of wave packet, off-diagonal entries of $Y$ becomes zero. 
We can regard this process as a gradual Higgsing, 
\begin{align}
{\rm SU}(N)
\to
{\rm SU}(N-1)\times{\rm U}(1)
\to
\cdots
\to
{\rm SU}(M)\times{\rm U}(1)^{N-M}
\to
\cdots
\to
{\rm U}(1)^{N-1}. 
\end{align}
When D0-branes emitted from black hole go very far, weakly-coupled gravity is not a good description. Still, some important features of black hole evaporation are captured, as we will see below.  

Let regions $A$, $B$ and $C$ be black zero-brane, emitted D0-branes and off-diagonal entries connecting them (Fig.~\ref{fig:IIA-D0-evaporation}, top row). 
Off-diagonal entries can be treated perturbatively, as we did in a sample calculation in Sec.~\ref{sec:Bekenstein-Hawking}. 
Qualitatively, the von Neumann entropy behaves as shown in Fig.~\ref{fig:IIA-D0-evaporation}.

Off-diagonal elements connecting the black zero-brane and emitted D0-branes (region $C$) are completely frozen to the ground state due to Higgsing. 
In the case of small black hole described by 4d SYM, a large number of degrees of freedom are frozen due to confinement. 
Either way, those `frozen' degrees of freedom (region $C$) are strongly entangled with `radiations' (region $B$) at late time. 

In this example, $S_A=S_{B\cup C}$ starts from zero, increases at first and then decreases to zero, resembling the Page curve. 
On the other hand, $S_B$ starts from zero, monotonically increases and saturates at finite value when the `evaporation' ends. 

Note also that the decoupling of off-diagonal degrees of freedom leads to a negative specific heat~\cite{Berkowitz:2016znt,Berkowitz:2016muc}. 
Because the energy per dynamical degrees of freedom increases, temperature goes up. 
\section{Conclusion, discussion and outlook}\label{sec:discussion}
In this paper, we introduced the notion of matrix entanglement for large-$N$ gauge theories, including matrix model and QFT. 
Based on previous references, we argued that matrix degrees of freedom naturally split into multiple sectors that have natural geometric interpretations in terms of string theory. 
When applied to fuzzy-sphere states in the BMN matrix model, matrix entanglement can be used to define the spatial entanglement in emergent QFTs nonperturbatively.  
We also considered the evaporation of small black hole and pointed out that the confined degrees of freedom play important role in the derivation of the Page curve. 
The confined sector behaves similarly to the entanglement island in several ways. Currently, we do not know whether the confined sector is the gauge-theory counterpart of the entanglement island, or they are actually different and the confined sector gives the Page curve without involving the island. 

To conclude the paper, let us discuss several other open problems.

\subsubsection*{Interior or exterior?}
\begin{figure}[htbp]
  \begin{center}
   \includegraphics[width=100mm]{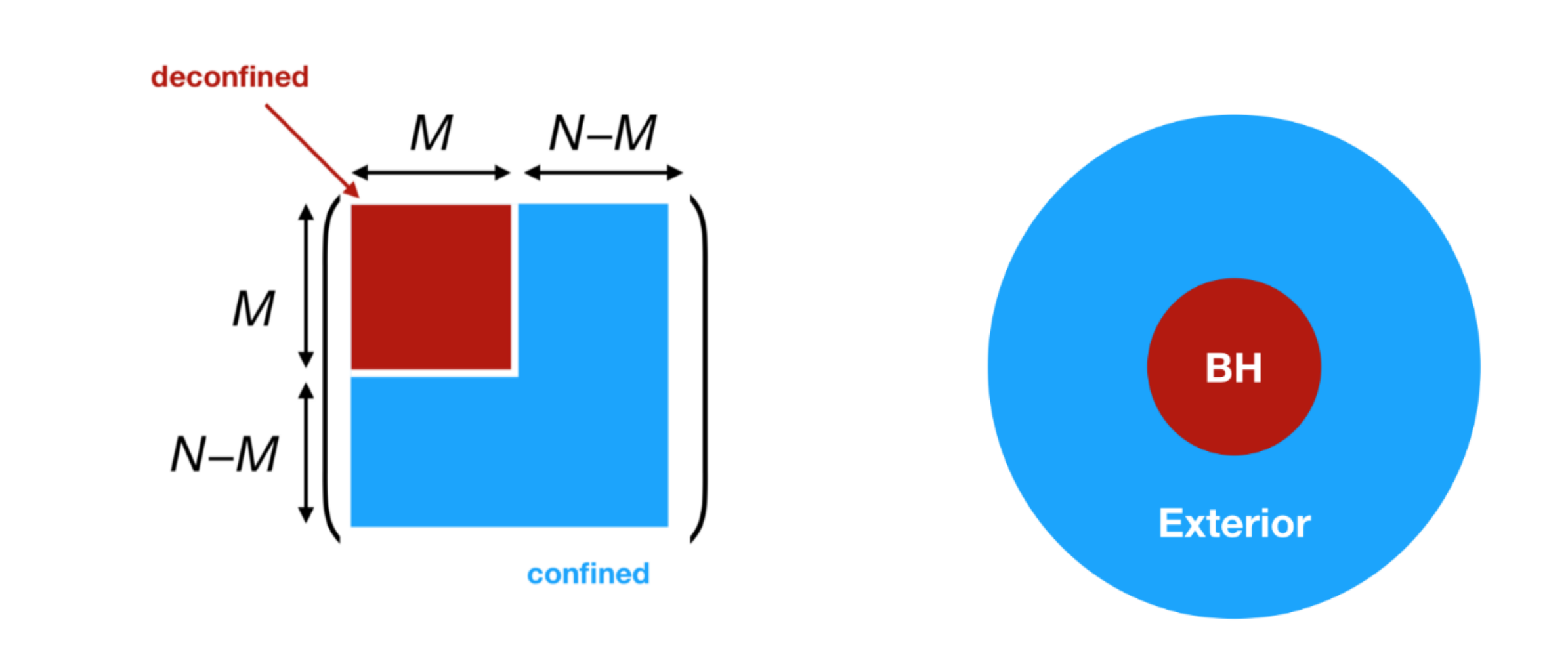}
  \end{center}
  \caption{
Fig.~4 in Ref.~\cite{Hanada:2019czd}. Is it consistent with the interpretation provided in this paper?
  }\label{fig:Figure-from-anatomy-of-deconfinement}
\end{figure}

In this paper, we discussed a possibility that the confined degrees of freedom in the partially-deconfined states correspond to the entanglement island sitting behind the horizon.
On the other hand, in the past, three of the authors (MH, AJ, CP) and collaborators speculated that the confined sector describes the exterior; see e.g., Fig.~\ref{fig:Figure-from-anatomy-of-deconfinement}, which is taken from Ref.~\cite{Hanada:2019czd}. What would be the precise interpretation then?

The speculation in Ref.~\cite{Hanada:2019czd} was partly based on the assumption that the `eigenvalues' of matrices are spread out even at low energy~\cite{Polchinski:1999br,Susskind:1998vk} and fill the significant portion of bulk geometry (for the case of AdS$_5$/CFT$_4$, the distance of order of AdS radius from the center of the bulk). This assumption is not necessarily valid for low-energy states, because `eigenvalues' used there describe the location of D-branes only near the boundary of the bulk~\cite{Hanada:2021ipb,Hanada:2021swb}. 
The eigenvalues of $Y_I$ (center of wave packet) can provide us with a better description which is valid even near the center of the bulk,   
and by using $Y_I$ we can show that most D-branes are well localized in the low-energy states. 
In particular, D-branes in the confined sector sit at the origin of the bulk.
When a small black hole (deconfined sector) is formed at the center of the bulk, those D-branes in the confined sector sit behind the horizon. In this sense the confined sector describes the interior. Still, the confined sector plays a crucial role for the exterior: the bulk geometry asymptotically coincides with black $p$-brane (AdS$_5\times$S$^5$ in the case of $p=3$) because of the existence of the confined sector, and the curvature is determined by the total number of degrees of freedom that includes both confined and deconfined sectors.
A larger fraction of the bulk is covered by black hole when the confined sector is smaller. 
Furthermore, it is possible to excite some degrees of freedom from the confined sector and probe exterior geometry.
Therefore, we are tempted to think that the confined sector describes both interior and exterior.  
Note also that, when the AdS radius is very large and black hole is very small, black hole can sit away from the origin, at least for some time, then the D-branes in the confined sector are not necessarily behind the horizon of this black hole. (They can be still regarded as sitting behind the horizon of supersymmetric black hole whose horizon area is zero.) 
It appears that the entire black hole geometry -- both interior and exterior -- is encoded in the way matrix degrees of freedom interact with each other. 

\subsubsection*{More on target-space entanglement}

In the target-space entanglement approach proposed in the past~\cite{Das:2020jhy,Das:2020xoa,Hampapura:2020hfg}, $X_1$ was diagonalized. 
Instead we can diagonalize $Y_1$ and define the separation of matrix degrees of freedom.
If the eigenvalues of $Y_1$ are sufficiently separated, these two approaches give the same result. However, in the strongly coupled region where the separation is not large, matrix entanglement captures emergent geometry more precisely. 
More generally, it would be better to think that $Y_I$'s describe matrix-valued target space, because the block-diagonal structure is of crucial importance. 

In Ref.~\cite{Anous:2019rqb}, extremal surfaces in the black zero-brane geometry dual to the D0-brane matrix model have been investigated. It would be interesting to study if these extremal surfaces correspond to some partition of matrix degrees of freedom. 
\subsubsection*{What are the geometry made of?}
We conjecture that the area of various extremal surfaces correspond to entanglement entropy associated with physically meaningful partition of matrix degrees of freedom. If this is correct, then the information regarding the metric in the gravitational geometry can be obtained from quantum entanglement, by using the duality to define quantum gravity.
This approach fits into the direction of study called `It from qubit'. 

Another natural way to determine the gravitational geometry is to see the interactions. For example, if one of the matrix degrees of freedom is used as a probe D-brane, its motion should be described by the Dirac-Born-Infeld action on a geometry sourced by other matrix degrees of freedom~\cite{Maldacena:1997re}.\footnote{
In the past, this approach was thought to be subtle because it was believed that it is difficult to relate the location of the probe to `matrix eigenvalues'. As we have seen, this issue has been resolved~\cite{Hanada:2021ipb}. 
} 
This approach seems to be practically useful at least outside the horizon. 

Although it is natural to expect that these two approaches define the same geometry, such an agreement is far from trivial. 
By confirming the consistency between these two approaches, we might be able to provide strong tests of the validity of the emergent geometry in holography. 
For explicit confirmation, machine-learning-motivated variational Monte Carlo method~\cite{Han:2019wue} and quantum simulation (including quantum-classical hybrid algorithm which can run on NISQ devices)~\cite{Preskill:2018fag} may be useful.

\subsubsection*{Numerical simulations}
To study small black hole quantitatively from QFT side, numerical methods will be needed.
4d SYM may be difficult at this moment, but the M-theory parameter region of D0-brane matrix model is within reach~\cite{Bergner:2021goh}.
A tricky issue is that the small black hole corresponds to the maximum of free energy in the canonical ensemble, 
which is not efficiently sampled by standard lattice simulation methods. 
This problem can be circumvented rather easily by constraining the value of the Polyakov loop~\cite{Bergner:2019rca}. 
A gauge fixing in which confined and deconfined sectors are clearly separated can be achieved by combining the static-diagonal gauge and appropriate permutations of Polyakov line phases~\cite{Watanabe:2020ufk}. 
Once the gauge is fixed properly, the R\'{e}nyi entropy can be calculated by using the Replica trick. 
It would be nice if we could reproduce black hole entropy quantitatively.

Other potentially useful numerical approaches include machine-learning-motivated variational Monte Carlo method and quantum simulation~\cite{Han:2019wue,Preskill:2018fag,Rinaldi:2021jbg}. 
   
\subsubsection*{Other theories, other matter contents}
In this paper, we considered only the matrix-valued fields. It is straightforward to generalize the construction to other representations, i.e., theories with fundamental fields such as the O($N$) vector model or QCD. 
It might be interesting to study the entanglement between different component of the vectors in the O($N$) vector model with singlet constraint, because this model has dual higher spin theory on AdS~\cite{Klebanov:2002ja}, and the mapping between the two sides of the duality is understood~\cite{deMelloKoch:2018ivk,Aharony:2020omh} better than in the duality between super Yang-Mills theory and superstring theory.  
Note that partial deconfinement takes place in the O($N$) vector model as well~\cite{Hanada:2019czd}, which provides us with a natural way of splitting vector components into two pieces. See also~\cite{Jevicki:2021ddf} for the construction of the thermofield double states. 

The SYK model does not have gauge symmetry and hence the definition of entanglement might be conceptually simpler. 
For the calculation of entanglement associated with the partition of fermions to two groups, see e.g.,~\cite{Fu:2016yrv,Zhang:2020kia}. 
The coupled-SYK model~\cite{Maldacena:2018lmt} has a phase structure analogous to 4d maximal super Yang-Mills on three-sphere, in that a phase with negative specific heat resembling a small black hole exists. Perhaps there is a natural splitting of degrees of freedom similar to partial deconfinement: in the ‘confined’ sector fermions in the left and right copies would be entangled, while in the ‘deconfined’ sector the entanglement would be lost. Such a splitting, if exist, can explain the dual gravitational interpretation proposed in~\cite{Maldacena:2018lmt}: the traversable wormhole (confined sector) is gradually destroyed as black hole (deconfined sector) grows~\cite{Alet:2020ehp}.  Such a splitting can be confirmed in toy models such as the coupled-matrix model~\cite{Alet:2020ehp}.

\subsubsection*{Definition of matrix entanglement without relying on gauge fixing}
In this paper, we defined the matrix entanglement after gauge fixing. Such a definition is natural at large $N$ and for the class of states we considered, because of the emergence of superselection sectors.  
To consider other kinds of states or finite-$N$ effects, a definition which does not rely on gauge fixing may be needed. See Ref.~\cite{Das:2020xoa} for the gauge-invariant construction of the target-space entanglement entropy.
\begin{center}
\section*{Acknowledgement}
\end{center}
The authors would like to thank Sumit Das, Alexander Frenkel, Chethan Krishnan, Raghu Mahajan, Jonathan Maltz, Gautam Mandal, Suvrat Raju, Enrico Rinaldi, Steve Shenker, Bo Sundborg, Brian Swingle, Sandip Trivedi, and Benson Way. 
VG thanks STFC for Doctoral Training Programme funding (ST/W507854-2021 Maths DTP). 
MH was supported by the STFC Ernest Rutherford Grant ST/R003599/1.
CP is supported by the Fundamental Research Funds for the Central Universities, by funds from the University of Chinese Academy of Science, by funds from KITS and also funds from NSFC No.~12175237.
\vspace{5mm}
\begin{center}
{\Large \textbf{Data management}}
\end{center}
No additional research data beyond the data presented and cited in this work are needed to validate the research findings in this work.
\appendix
\section{From canonical quantization to path integral
}\label{sec:H-to-L}
\hspace{0.51cm}
In this appendix, we show how the partition function \eqref{eq:Z-H-ext-MM} expressed in terms of the extended Hilbert space is related to path integral. Here we consider matrix quantum mechanics. Generalization to QFT is straightforward. We take the gauge group $G$ to be U($N$). 
In a rather trivial manner, the thermal partition function \eqref{eq:Z-H-ext-MM} can be rewritten as 
\begin{align}
Z(T)
&=
\frac{1}{[{\rm volU}(N)]^K}
\int\left(\prod_{k=1}^KdU_k\right)
{\rm Tr}_{{\cal H}_{\rm ext}}
\Bigl(
\hat{U}_{(K)}
e^{-\frac{H(\hat{P},\hat{X})}{TK}}
\hat{U}_{(K-1)}^{-1}\hat{U}_{(K-1)}
\nonumber\\
&
\quad
e^{-\frac{H(\hat{P},\hat{X})}{TK}}
\hat{U}_{(K-2)}^{-1}\hat{U}_{(K-2)}
\cdots
\hat{U}_{(1)}^{-1}\hat{U}_{(1)}
e^{-\frac{H(\hat{P},\hat{X})}{TK}}
\Bigl)
\nonumber\\
&=
\frac{1}{[{\rm volU}(N)]^K}
\int\left(\prod_{k=1}^KdU_{(k)}\right)
\int\left(\prod_{k=1}^KdX_{(k)}\right)
\nonumber\\
&
\qquad
\langle X_{(K)}|
\hat{U}_{(K)}
e^{-\frac{H(\hat{P},\hat{X})}{TK}}
\hat{U}_{(K-1)}^{-1}
|X_{(K-1)}\rangle
\nonumber\\
&
\qquad
\times
\langle X_{(K-1)}|
\hat{U}_{(K-1)}
e^{-\frac{H(\hat{P},\hat{X})}{TK}}
\hat{U}_{(K-2)}^{-1}
|X_{(K-2)}\rangle
\nonumber\\
&
\qquad
\times
\cdots
\times
\langle X_{(1)}|
\hat{U}_{(1)}
e^{-\frac{H(\hat{P},\hat{X})}{TK}}
|X_{(K)}\rangle. 
\end{align}
Terms on the right hand side can be expressed by using the Lagrangian as 
\begin{align}
&
\langle X_{(k)}|
\hat{U}_{(k)}
e^{-\frac{H(\hat{P},\hat{X})}{TK}}
\hat{U}_{(k-1)}^{-1}
|X_{(k-1)}\rangle
\nonumber\\
&\quad=
\langle U_{(k)}^{-1}X_{(k)}U_{(k)}|
e^{-\frac{H(\hat{P},\hat{X})}{TK}}
|U_{(k-1)}^{-1}X_{(k-1)}U_{(k-1)}\rangle
\nonumber\\
&\quad=
\int dP
\langle U_{(k)}^{-1}X_{(k)}U_{(k)}|
e^{-\frac{H(\hat{P},\hat{X})}{TK}}
|P\rangle\langle P|
U_{(k-1)}^{-1}X_{(k-1)}U_{(k-1)}\rangle
\nonumber\\
&\quad=
\int dP
e^{i{\rm Tr}[P(U_{(k)}^{-1}X_{(k)}U_{(k)}-U_{(k-1)}^{-1}X_{(k-1)}U_{(k-1)})]}
e^{-H(P,U_{(k)}^{-1}X_{(k)}U_{(k)})/(TK)}
\nonumber\\
&\quad=
e^{-\frac{1}{2}KT{\rm Tr}[(U_{(k)}^{-1}X_{(k)}U_{(k)}-U_{(k-1)}^{-1}X_{(k-1)}U_{(k-1)})^2]}
e^{-V(U_{(k)}^{-1}X_{(k)}U_{(k)})/(TK)}
\nonumber\\
&\quad\simeq
e^{-L[D_t(U_{(k)}^{-1}X_{(k)}U_{(k)}),(U_{(k)}^{-1}X_{(k)}U_{(k)})]/(TK)}
\nonumber\\
&\quad=
e^{-L[D_tX_{(k)},X_{(k)})]/(TK)}. 
\end{align}
Here we used 
\begin{align}
U_{(k-1)}U_{(k)}^{-1}\equiv e^{iA_{(k)}/(KT)}
\end{align}
and
\begin{align}
X_{(k)}-(U_{(k-1)}U_{(k)}^{-1})^{-1}X_{(k-1)}(U_{(k-1)}U_{(k)}^{-1})
\simeq
\frac{D_tX_{(k)}}{KT}.   
\end{align}
By taking $K\to\infty$ limit, we obtain the standard expression in the path-integral formalism:
\begin{align}
Z(T)
&=
\int [dA] [dX]e^{-\int dt L[D_tX,X]}. 
\end{align}
The same derivation applies to real-time correlation functions. 
\bibliographystyle{utphys}
\bibliography{matrix-entanglement-hepth-v2}

\end{document}